\documentclass[aps,prd,preprintnumbers,nofootinbib,onecolumn,superscriptaddress]{revtex4}
\usepackage[dvips,dvipdfmx]{graphicx}
\usepackage{bm,latexsym,amsmath,amssymb,amsfonts,mathrsfs,calc}
\usepackage[final]{pdfpages}
\usepackage{comment}
\usepackage{xcolor}

\usepackage[colorlinks,linkcolor=blue,anchorcolor=violet,citecolor=blue]{hyperref}

\begin{document}

\newcommand{\simgt}{\lower.5ex\hbox{$\; \buildrel > \over \sim \;$}}
\newcommand{\simlt}{\lower.5ex\hbox{$\; \buildrel < \over \sim \;$}}
\newcommand{\nan}[1]{\textcolor{blue}{#1}}

\def\D{{\cal D}}
\def\msq{{m_{\phi}^2}}
\def\mp{{m_{\phi}}}
\newcommand*{\diff}{\mathop{}\!\mathrm{d}}

\vspace{2cm}  

\begin{flushright}
\end{flushright}

\title{\large Dark energy model with very large-scale inhomogeneity}

\author{Yue Nan}
\affiliation{
   Department of Physics, Graduate School of Science, Hiroshima University,
   Higashi-Hiroshima 739-8526, Japan}

\author{Kazuhiro Yamamoto}
 \affiliation{
  Department of Physics, Kyushu University, 744 Motooka, Nishi-Ku, Fukuoka 819-0395, Japan}
  \affiliation{Research Center for Advanced Particle Physics, Kyushu University, 744 Motooka, Nishi-ku, 
 Fukuoka 819-0395, Japan}




\begin{abstract}
  We consider a dynamical model for dark energy based on an ultralight mass scalar field with very large-scale inhomogeneities. 
  This model may cause observable impacts on the anisotropic properties of the cosmic microwave background (CMB) intensity and luminosity distance.
  We formulate the model as the cosmological perturbations of the superhorizon scales, focusing on the local region of our universe. Moreover, we investigated the characteristic properties of the late-time evolution of inhomogeneous dark energy. Our numerical solutions show that the model can mimic the standard $\Lambda$CDM cosmology while including spatially dependent dark energy with flexible ranges of the model parameters. 
  We put a constraint on the amplitude of these inhomogeneities of the dark energy on very large scales with the observations of the CMB anisotropies. We also discuss their influence on the estimation of the luminosity distance.
\end{abstract}

\maketitle

\def\bar{\overline}
\def\x{{z}}
\def\L{L}
\def\bmx{{\bm x}}
\def\bmy{{\bm y}}
\def\bmp{{\bm p}}
\def\bmk{{\bm k}}
\def\deltatau{{\Delta\tau}}
\def\barphi{{\bar\phi}}
\def\barpi{{\pi_\chi}}
\def\calh{\mathcal{H}}
\section{Introduction}
\label{sec:intro}
The observations of the redshifts of the distant Type Ia supernovae (SNe Ia) imply the existence of an unknown repulsive interaction that
accelerates the expansion of the universe in relatively late times~\cite{sne0,sne1,sne2,sne3,sne4}; otherwise, this fact suggests the breakdown of 
general relativity on cosmological scales. 
  To account for these observations, dark energy has become an essential component of cosmology since the late 1990s, in addition to the cold dark matter (CDM)~\cite{Peebles2002}. 
  Albert Einstein first introduced a cosmological constant $\Lambda$ into general relativity to establish a static universe. 
Since the expansion of the universe was discovered and the big-bang cosmology became a paradigm after the discovery of the cosmic microwave background (CMB), the cosmological constant $\Lambda$ was revived and discussed occasionally (see, Ref.~\cite{SWein1989} for a review).
The cosmological constant $\Lambda$ is now included in the standard model of cosmology as the simplest model of dark energy to explain the accelerated expansion, which has also been tested by observing the CMB and baryon acoustic oscillations (BAO) in the context of structure formation theories~\cite{Weinberg}.
The scenario is summarized as the well-known standard $\Lambda$CDM cosmological model~\cite{Planck2018I, Planck2018VI}, where dark energy consists of approximately 70\% of the total energy density of the universe at the present epoch. 

Many dark energy models have been proposed as variants of the cosmological constant, where the equation of state (EoS) $\omega$ is defined by the ratio of pressure $p$ to energy density $\rho$ as $\omega=p/\rho$ is a typical quantity used to characterize the property of the dark energy. Early observations of SNe Ia 
constrained that $\omega < -1/3$ for dark energy, 
which have been followed by more precise observations, suggesting that the dark energy EoS would be very close to the cosmological constant, with $\omega=-1$. 
As the energy density of radiation $\rho_r$ and matter $\rho_m$ decays with  $\rho_r\propto a^{-4}$, $\rho_m\propto a^{-3}$  with the scale factor of the universe $a$, and dark energy seems to behave as an almost constant 
and homogeneous background of the universe, its energy density is suggested to 
become dominant in the late times of the universe when $a\gtrsim0.5$. 
Hence, the property of dark energy is important for the evolution of the universe, especially in the late times and in the future. 

The large-scale structure of matter distributions serves as a useful probe for dark energy EoS because the BAO signature is useful as a standard ruler. 
Furthermore, the growth of clustering of the matter is affected by dark energy.
On the other hand, these also gave rise to another mysterious aspect of dark energy as a famous fine-tuning problem, i.e. ``cosmological constant problem'' (see Refs.~\cite{Peebles2002,SWein1989}). The problem is why the dark energy density is of the same order as the matter density at the present epoch, 
much smaller than the prediction from a naive expectation of modern particle physics theories, while its EoS implies linkage with the vacuum energy of quantum fields. 
These problems may be closely related to the origin and nature of dark energy, which remains to be explored.

Many theoretical models of dark energy have been investigated~\cite{Weinberg,Peebles2002},
in which dynamical models are very interesting~\cite{Tsujikawa:2013fta,Ringeval,
Glavan1,Glavan2,Glavan3,Glavan4,DEquantum,DEquantum2}, 
because they are related to the field theory associated with the primordial high-energy epoch of the universe and fundamental theories of theoretical physics~\cite{SWein1989,Peebles2002,Linder2020}. 
Particularly interesting ones are the dynamical models based 
on the quantum fluctuations of ultralight scalar fields~\cite{Ringeval,Glavan1,Glavan2,Glavan3,Glavan4,DEquantum,DEquantum2}, which reveal an interesting connection to the string axiverse scenario~\cite{Witten,Arvanitaki,Obied,Agrawal,Garg:2018reu,Ooguri,Garg2019,Visinelli:2018utg}. 
As in the $\Lambda$CDM model and in many models of dark energy,
a basic assumption of their property is spatial isotropy and homogeneity, which follows the cosmological principle. Nevertheless, since the late-time expansion of the universe is dominated by dark energy, 
some interesting outcomes may occur to affect cosmological observables if large-scale inhomogeneities of dark energy arise, which could be tested by various cosmological observations~\cite{ABM,TSJ,Jassal2010,Yamauchi2018,Linder2020}.

On the other hand, anomalous features in the CMB anisotropies have been pointed out 
by some authors \cite{cmb,fosalba}. Although
the cosmic variance limits the ability for our precise
comparison between theoretical predictions and observations, 
there is the possibility that the low CMB multipoles provide us with a clue for physics beyond the standard cosmological model for dark energy~\cite{Gordon2005,Polastri2015}.
The general interpretation for the CMB dipole 
anisotropy is our peculiar motion toward a CMB rest frame, related to a dragging toward the great attractor in the sky; at least part of the peculiar motions is interpreted as evidence of gravitational bounding~\cite{Tully2008, Courtois2013}. 
The latest result shows the validity of the interpretation of the CMB dipole by the peculiar motion \cite{Ferreira}.
However, the result does not necessarily mean that all of the CMB dipole anisotropy could be entirely explained by the canonical scenario of peculiar motion~\cite{PlanckLVI,Secrest}. We will present a dark energy model with very large-scale inhomogeneities, including an intrinsic dipole component as a possible solution.

Recently, the Hubble tension problem has also garnered attention due to the precision of the observations.
The present expansion rate $H_0$ locally measured from standard candles, such as SNe Ia~\cite{Riess2011} 
and that inferred from the BAO statistics on CMB fluctuations~\cite{Bennett2013, Planck2018VI},
has shown nontrivial deviations from each other. Many attempts have been made to ease or explain this tension, 
and among them still stands out the possibility that this tension is related to new physics concerning dark energy 
beyond the standard $\Lambda$CDM model~\cite{Mortsell2018}.
Recent investigations using scaling relations of galaxy clusters in Refs.~\cite{Migkas2020,Migkas2021} reported that the variation in luminosity distance $d_L$ appears to exist in different regions of the sky, potentially suggesting anisotropy in the local expansion rate $H_0$, whose line was followed by Refs.~\cite{Eoin1,Eoin2}. Some correlation of $H_0$ anomalies with the CMB dipole direction is commonly implied by these works, which also facilitates the motivation of our work.

To shed light on the problems concerning dark energy, the authors investigated a model for dark energy 
with large-scale stochastic fluctuations assumed in an open universe associated with a specific inflationary scenario~\cite{scmde1}. 
These fluctuations will be translated into large-scale spatial inhomogeneities and time-dependent dark energy EoS in the evolution of the universe.
In the present work, we consider a general dynamical model for dark energy with large-scale spatial inhomogeneities consisting of a scalar field $\phi$ by handling them in the framework of the cosmological perturbation theory. This model may introduce 
some observable effects on the anisotropies of the cosmological observations to address the problems concerning the dark energy property mentioned previously.

The remainder of this paper is organized as follows. In Sec.~\ref{sec:basic}, we propose a basic formulation for the model and its cosmological setups. Then, we use the formulation to derive the Einstein equations for the system as well as the equations of motion: 
for both the dark energy represented by the dynamical scalar field $\phi$ and the matter component in the late-time universe. In Sec.~\ref{sec:numer}, 
we use the analytic approximations to solve for the equations in the limit $a \ll 1$, where $a$ denotes the scale factor of the universe. 
This is useful to determine the necessary initial conditions for the numerical solution to the late-time cosmological 
evolution of the system. 
In Sec.~\ref{sec:appli}, we consider the possible effects of large-scale dark energy fluctuations on cosmological observations, such as the CMB temperature power spectra and luminosity distance. Section~\ref{sec:discuss} is devoted to summarizing our results and brief discussions.
The appendices provide additional explanations for specific technical details. 
In Appendix~\ref{appen:matrix}, explicit forms of the matrices used in the definition of the perturbations are presented, and their relations with multipole expansion are discussed. 
In Appendix~\ref{appen:fluideq}, we show the consistency of the derived equations with previous works~\cite{scmde1,WHuthesis}, especially for the superhorizon Euler equation of the matter component.
Appendix~\ref{appen:background} provides additional details for the background solutions and the analytic approximations.
Appendix~\ref{appen:EOSCPL} shows the dark energy EoS in our model and its relation to the
Chevallier-Polarski-Linder (CPL)
parametrization~\cite{ChePolar,Linde0}. In Appendix~\ref{appen:ld}, we show that our application of the model to the correction of the luminosity distance is valid and consistent with previous works~\cite{FS1989,sasaki1987}. Finally, in Appendix~\ref{appen:transf}, we present a helpful toolkit for transforming equations between forms with respect to different variables in our model.

\section{Basic Formulation}
\label{sec:basic}
In the present paper, 
motivated by a previous model with supercurvature-mode dark energy associated with an open universe scenario~\cite{DEquantum,DEquantum2,Aoki},  
we consider the evolution of dark energy with superhorizon large-scale inhomogeneities and its possible imprints on cosmological observations by characterizing the inhomogeneities analytically.
To formulate these inhomogeneities, we start with following the cosmological setup of metric perturbations. 

\subsection{Fundamental setups}
\label{sec:setup}
The characteristic feature of the dark energy model 
previously proposed in Refs.~\cite{scmde1,Aoki} is 
the spatial inhomogeneities of the dark energy density
on the very large scales. 
Following the scenario, such large-scale spatial inhomogeneities of
dark energy originated from the vacuum fluctuations of the supercurvature modes of a scalar field during an open inflationary scenario~\cite{Aoki,Yamauchi2011}.
An ultralight scalar field $\phi$ with spatial fluctuations 
taking nonlinear amplitude on the supercurvature scales is responsible for 
the dark energy in the scenario.
Because the horizon size of our universe is much smaller than 
the scales of the inhomogeneities of the dark energy, the 
breaking of the cosmological principle is small within the
observable universe, which might enable us to escape from the
observational constraints. 

In the present paper, we formulate a phenomenological model of dark energy that slightly breaks the cosmological principle by mimicking
the previous model~\cite{scmde1,Aoki}. 
We consider a dark energy model of a scalar field 
spatially varying on the superhorizon scales 
on the spatially flat background universe, for simplicity, by assuming
\begin{align}
  d s^{2}
  =a^{2}(\eta)\left[-(1+2 \Psi) d \eta^{2}+(1+2 \Phi) \delta_{i j} d x^{i} d x^{j}\right],
  \label{eq:metric}
\end{align}
where $\delta_{i j}$ is the Kronecker delta $\delta_{i j}$, $a(\eta)$ is the scale factor of the universe with the 
conformal time $\eta$, 
and $\Psi$ and $\Phi$ are the metric perturbations that we want to characterize later.

Now, we set the cosmological metric perturbation as $\Psi$, considering only the large-scale superhorizon mode perturbations. 
In Ref.~\cite{scmde1}, it was discussed that the inhomogeneities induced by superhorizon fluctuations are dominated by dipole and quadrupole components among all possible contributions. Now, neglecting higher multipoles, we can explicitly write out the metric perturbations as
\begin{align}
  \Psi=\epsilon_1\sum^3_{m=1}\Psi_{1(m)}(\eta)P_i^{(m)}x^{i}+\epsilon_2\sum^5_{m=1}\Psi_{2(m)}(\eta)P^{(m)}_{ij} x^i x^j,
  \label{def:Psi}
  \\ 
  \Phi=\epsilon_1\sum^3_{m=1}\Phi_{1(m)}(\eta)P_i^{(m)}x^{i}+\epsilon_2\sum^5_{m=1}\Phi_{2(m)}(\eta)P^{(m)}_{ij} x^i x^j,
  \label{def:Phi}
  \\
  \phi=\phi_0(\eta)+\epsilon_1\sum^3_{m=1}\phi_{1(m)}(\eta)P_i^{(m)}x^{i}+\epsilon_2\sum^5_{m=1}\phi_{2(m)}(\eta)P^{(m)}_{ij} x^i x^j,
  \label{def:field}
\end{align}
where $P^{(m)}_{i}$ and $P^{(m)}_{ij}$ are the vectors of traceless  matrices related to the multipole expansion of the perturbations to the spatial basis, whose expressions are explicitly given in Appendix~\ref{appen:matrix}.
We use $\phi$ to denote the ultralight scalar field we assume as the source of dark energy with large-scale spatial inhomogeneities. Here $\epsilon_1$ and $\epsilon_2$ are introduced to explicitly express 
the order of perturbations for the dipole and the quadrupole, 
which can be included in the perturbations. We set 
$\epsilon_1$ and $\epsilon_2$ to be unity later.
Considering a standard CDM scenario, we can write the perturbations 
for the matter density distribution as
\begin{align}
  &\rho=\rho_{0}(\eta)+\epsilon_1\sum^3_{m=1}\rho_{1(m)}(\eta)P_i^{(m)}x^{i}+\epsilon_2\sum^5_{m=1}\rho_{2(m)}(\eta)P^{(m)}_{ij} x^i x^j,
  \label{def:den}
  \end{align}
  and we define the velocity field as
 \begin{align}
  &u_i\equiv \partial_i \overline V,
  \label{def:velo}
\end{align}
with constraints $u_{\mu}u^{\mu}=-1$, where
$\bar V$ is the velocity potential, which is expressed as
\begin{align}
  &\overline V=\epsilon_1\sum^3_{m=1}V_{1(m)}(\eta)P_i^{(m)}x^{i}+\epsilon_2\sum^5_{m=1}V_{2(m)}(\eta)P^{(m)}_{ij} x^i x^j.
\end{align} 
Here $\Psi_{\ell(m)}$, $\Phi_{\ell(m)}$, $\phi_{\ell(m)}$, 
$\rho_{\ell(m)}$, $V_{\ell(m)}$ with $\ell=1,2$ are the 
coefficients of the dipole and the quadrupole components, 
and $\phi_0$ and $\rho_0$ are the background quantities.

\subsection{Essence of the equations}
The evolution of the system is described by the Einstein equations 
\begin{align}
  G^{\mu}_{}{}_{\nu}=8 \pi G\left(T^{(\phi) \mu}_{}{}_{\nu}+T^{(\rm M)}{}^{\mu}_{}{}_{\nu}\right), 
  \label{eins}
\end{align}
with the energy momentum tensors for the scalar field with mass $m$ and the 
matter component,
\begin{align}
  &T_{\mu \nu}^{(\phi)}=\partial_{\mu} \phi \partial_{\nu} \phi-
  g_{\mu \nu}\left(\frac{1}{2} g^{\alpha \beta} \partial_{\alpha} \phi \partial_{\beta} \phi+\frac{1}{2} \msq \phi^{2}\right),
 ~~~~~~~~~
 T^{(\rm M)}_{\mu \nu}=\rho u_{\mu} u_{\nu},
\end{align}
and the equations of motion for the scalar field $\phi$ 
and the conservation law for the matter component,
\begin{align}
  \frac{1}{\sqrt{-g}} \partial_{\mu}\left(\sqrt{-g} g^{\mu \nu} \partial_{\nu} \phi\right)-\msq \phi=0,
  \label{eomscm}
\end{align}
\begin{align}
  \nabla_{\mu} T^{\rm (M)}{}^{\mu}_{}{}_{\nu}=0.
  \label{eomdm}
\end{align}

The EoS of the dark energy field $\phi$ is an important quantity characterizing its properties and evolution. 
From the standard formula for the energy density and the pressure
of a scalar field, taking the form of a scalar field potential $V(\phi)=\msq\phi^2/2$ into account
we obtain the equation of state $\omega_\phi$ as 
\begin{align}
  \omega_\phi 
  \equiv \frac{P_\phi}{\rho_\phi}
\simeq
  -\frac{2a^2 V(\phi)-\dot\phi^2}{2a^2 V(\phi)+\dot\phi^2}
 =-\frac{\msq a^2 \phi^2-\dot\phi^2}{\msq a^2\phi^2+\dot\phi^2},
  \label{def:eos}
\end{align}
where the dot denotes the differentiation with respect to the conformal time $\eta$.
Here we neglected the contribution from the spatial variations, 
which is small in our case.
The EoS depends on the dynamical evolution of $\phi$ and is a concordant generalization to the CPL parametrization (see Appendix~\ref{appen:EOSCPL})~\cite{ChePolar,Linde0}.

The linear expansion of Eqs.~(\ref{def:Psi})---(\ref{def:field}) ensures that Eqs.~(\ref{eins})---(\ref{eomdm}) give the same form as 
equations for each multipole component with indices $\ell=1,2$ and $m=1,2,3,4,5$. 
Indeed, the components with different $\ell$ indices, for example, $\Psi_{\ell=1}$ and $\Psi_{\ell=2}$, 
have different dimensions to the order of length by definition. 
Keeping this fact in mind, 
for simplicity of the notations, we neglect indices $(m)$ in the following parts, 
and use only the lower indices $\ell$ to denote the multipole components of these perturbations. In the following parts,
 we use lower indices $0$ for the background quantities and $\ell$ for the perturbations on the superhorizon scales. 

Using the conformal Hubble parameter $\calh=aH(a)=\dot a/a$ instead of Hubble parameter $H(a)$, Eq.~(\ref{eomscm}) yields: 
\begin{gather}
   \ddot\phi_0+2\calh\dot\phi_0+\msq a^2\phi_0=0,
  \\
   \ddot\phi_\ell+2\calh\dot\phi_\ell+\msq a^2\phi_\ell
  +\dot\phi_0(3\dot\Phi_\ell-\dot\Psi_\ell-4\calh\Psi_\ell)-2\ddot\phi_0\Psi_\ell =0.
\end{gather}
On the other hand, Eq.~(\ref{eomdm}) leads to
\begin{gather}
  3\calh\rho_0+\dot\rho_0=0,
  \label{eq:density0}
  \\
   3\calh\rho_\ell + \dot\rho_\ell + 3\rho_0\dot\Phi_\ell=0,
  \label{eq:density1}
  \\
   \dot V_\ell-a\Psi_\ell=0.
   \label{eq:flu}
\end{gather}
By defining the density perturbation as $\rho_\ell \equiv \rho_0 \delta_\ell$, it is obvious that Eqs.~(\ref{eq:density0}) and (\ref{eq:density1}), 
are consistent with those obtained from the continuity equation, and Eq.~(25) in  Ref.~\cite{scmde1} at a large-scale limit.
It is worth mentioning that the velocity equation in Eq.~(\ref{eq:flu}) is also consistent with Eq.~(26) in Ref.~\cite{scmde1}, which is obtained from the Euler equation (see Appendix~\ref{appen:fluideq}).

Defining $M_{\rm pl}^{-2}\equiv 8 \pi G$ for short, the Einstein equations can be written as
\begin{gather}
    -3 \calh^2 + M_{\rm pl}^{-2}(\frac{1}{2}\msq a^2\phi_0^2+\frac{1}{2}\dot\phi_0^2+a^2\rho_0)=0,
    \\
    \calh^2-2\frac{\ddot a}{a} +M_{\rm pl}^{-2}(\frac{1}{2}\msq a^2\phi_0^2-\frac{1}{2}\dot\phi_0^2)=0,
    \\
    -2(\calh\Psi_\ell-\dot\Phi_\ell) +M_{\rm pl}^{-2}(a \rho_0 V_\ell +  \dot\phi_0 \phi_\ell)=0,
    \\
    6 \calh (\calh\Psi_\ell-\dot\Phi_\ell) + M_{\rm pl}^{-2} \left(a^2\rho_\ell+\msq a^2 \phi_0 \phi_\ell 
    -\dot\phi_0(\dot\phi_0\Psi_\ell-\dot\phi_\ell) \right)=0,
    \\
    (2\frac{\ddot a}{a}-\calh^2)\Psi_\ell +\calh\dot\Psi_\ell -2\calh\dot\Phi_\ell -\ddot\Phi_\ell 
    +\frac{M_{\rm pl}^{-2}}{2}  \left(\msq a^2\phi_0\phi_\ell+ \dot\phi_0(\dot\phi_0\Psi_\ell-\dot\phi_\ell)\right)=0.
\end{gather}
We can classify these equations by the order of the perturbations, dividing them into the background equations that read
\begin{gather}
  \dot\rho_0+3\calh\rho_0=0,
  \label{eq:01}
\\
  \ddot\phi_0+2\calh\dot\phi_0+\msq a^2\phi_0=0,
  \label{eq:02}
  \\  
     -3 \calh^2 + M_{\rm pl}^{-2}(\frac{1}{2}\msq a^2\phi_0^2+\frac{1}{2}\dot\phi_0^2+a^2\rho_0)=0,
  \label{eq:03}
  \\
     \calh^2-2\frac{\ddot a}{a} +M_{\rm pl}^{-2}(\frac{1}{2}\msq a^2\phi_0^2-\frac{1}{2}\dot\phi_0^2)=0,
  \label{eq:04}   
\end{gather}
and first-order perturbative equations relying on the background as follows
\begin{gather}
  \dot\rho_\ell + 3\calh\rho_\ell +3\rho_0\dot\Phi_\ell=0,
  \label{eq:11}
  \\
   \ddot\phi_\ell+2\calh\dot\phi_\ell+\msq a^2\phi_\ell
  +\dot\phi_0(3\dot\Phi_\ell-\dot\Psi_\ell-4\calh\Psi_\ell)-2\ddot\phi_0\Psi_\ell =0,
  \label{eq:12}
  \\
   \dot V_\ell-a\Psi_\ell=0,
   \label{eq:13}
  \\
  -2(\calh\Psi_\ell-\dot\Phi_\ell) +M_{\rm pl}^{-2}(a \rho_0 V_\ell +  \dot\phi_0 \phi_\ell)=0,
  \label{eq:14}
  \\
   6 \calh (\calh\Psi_\ell-\dot\Phi_\ell) + M_{\rm pl}^{-2} \left(a^2\rho_\ell+\msq a^2 \phi_0 \phi_\ell 
    -\dot\phi_0(\dot\phi_0\Psi_\ell-\dot\phi_\ell) \right)=0,
  \label{eq:15}
  \\
   (2\frac{\ddot a}{a}-\calh^2)\Psi_\ell +\calh\dot\Psi_\ell -2\calh\dot\Phi_\ell -\ddot\Phi_\ell 
    +\frac{M_{\rm pl}^{-2}}{2}  \left(\msq a^2\phi_0\phi_\ell+ \dot\phi_0(\dot\phi_0\Psi_\ell-\dot\phi_\ell)\right)=0.
  \label{eq:16} 
\end{gather} 
After solving for the background, we can find out the evolution
of  large-scale perturbations originated from the fluctuations of 
the dark energy field $\phi$.

\section{Analytic Approximations and Numerical Solutions}
\label{sec:numer}
In this section, we consider solving the evolution equations 
obtained in the previous section for both the background and the perturbations. 
Because we are interested in the late-time evolution after the last scattering ($a_d\sim1/1100$), we first find the 
analytic approximates of the solutions based in 
the matter-dominant epoch, which are useful as the 
initial conditions for numerical evaluation when 
$a_d \simlt a \ll 1$.

\begin{figure}[b]
  \includegraphics[width=0.7\linewidth]{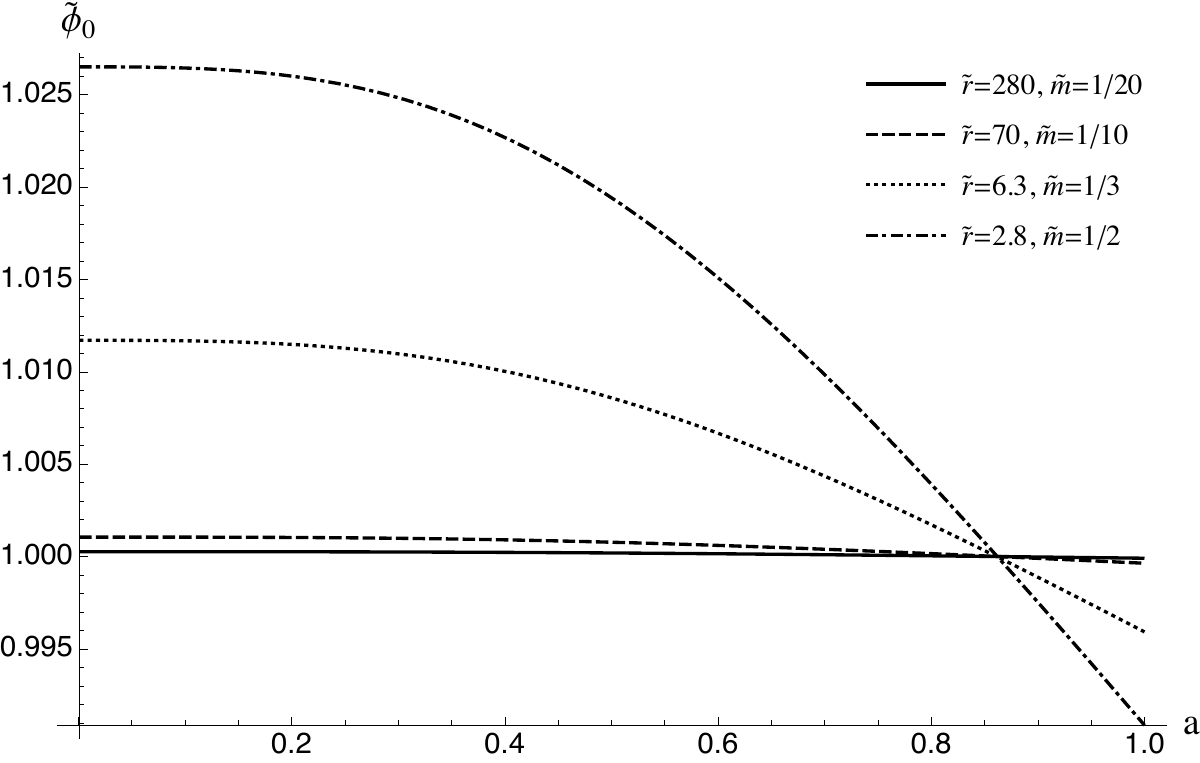}
  \caption{An example of the evolution of the background solutions $\widetilde{\phi}_0(a)$ as a function of the scale factor $a$ with the different sets of parameters for $\widetilde{r}$ and 
  $\widetilde{m}$ presented in the figure, which mimic $\Lambda$CDM universes 
  with $\Omega_m=0.3$ using Eq.~(\ref{eq:lcdmparameter}). 
  Notice that each model has different initial values for $\widetilde\phi_0$.
   We 
  observe from the figure that the lighter field $\phi$ is more ``frozen'' in its evolutionary history because $\widetilde{m}$ is normalized by the Hubble constant in Eq.~(\ref{def:ndm}). Here, the curve with $\widetilde{m}=1/20$ and $\widetilde{r}=280$ is most similar to 
  the cosmological constant model among the curves.}   
  \label{fig:phi0LCDM}
\end{figure}
\subsection{The background evolution}
\label{sec:numer_background}
First, we must solve the background evolution of our system in Eqs.~(\ref{eq:01})--(\ref{eq:04}) before considering the perturbations, 
which should yield cosmological observational constraints that 
models close to the $\Lambda$CDM models are favored. Moreover, we must use the 
observed value of the Hubble parameter at the present epoch to determine the dark energy density of the field $\phi$. 

Using these approximates we may infer the initial conditions of the background for numerical solutions of the background evolution.


To parametrize the equations, we introduce the cosmic time $t$ by $dt=ad\eta$. 
Defining tilde dimensionless quantities as
\begin{align}
  \tilde{t} &\equiv H_0 t,
  \label{def:ndt}
  \\
  \widetilde{\phi}_0 &\equiv {\phi_0 / \bar  \phi_0 } ,
  \label{def:ndphi}
  \\
  \widetilde r &\equiv \frac{1}{6}\left({\bar \phi_0 / M_{\rm pl}}\right)^2 ,
  \label{def:ndr}
  \\
  \widetilde{m} &\equiv \mp/H_0 ,
  \label{def:ndm}
  \\
 \widetilde{H} &\equiv H/H_0,
   \label{def:ndh}
\end{align}
we can obtain dimensionless ordinary differential equations using $\tilde{t}$ as independent variable as
\begin{align}
   \widetilde r \widetilde{m}^2\widetilde{\phi}_0^2(\tilde{t})+\widetilde r \left(\frac{\diff\widetilde{\phi}_0}{\diff\tilde{t}} \right)^2+\Omega_m a^{-3}
   &=\left(\frac{1}{a}\frac{\diff a}{\diff\tilde{t}}\right)^2,
  \label{eq:tnd1}
  \\
  \frac{\diff^2\widetilde{\phi}_0}{\diff\tilde{t}^2} + 3 \frac{1}{a} \frac{\diff a}{\diff\tilde{t}}
  \frac{\diff\widetilde{\phi}_0}{\diff\tilde{t}}+\widetilde{m}^2 \widetilde{\phi}_0
  &=0, 
  \label{eq:tnd2}
\end{align}
where $H_0$ is the Hubble constant, and 
$\bar\phi_0$ is a constant related to the initial  value of $\phi_0$.
If we use the scale factor $a$ instead of $t$, and use superscript $'$ to denote the derivative with respect to scale factor $a$, then the equations correspond to
  \begin{align}
    & \left(1-\widetilde{r} a^2  \widetilde{\phi}_0'^2\right)\widetilde{H}^2=\widetilde{r}\widetilde{m}^2\widetilde{\phi}_0^2+\Omega_m a^{-3},
    \label{eq:ha1}
    \\
    & 
      a^2 \widetilde{H}^2 \widetilde{\phi}_0''+  \left( 4 a \widetilde{H}^2 
 + a^2 \widetilde{H}\widetilde{H}' \right) \widetilde{\phi}_0'
   + \widetilde{m}^2\widetilde{\phi}_0=0. 
    \label{eq:phia2}
  \end{align}
Following Eq.~(\ref{eq:ha1}) we may also write out the dimensionless expansion rate as
  \begin{align}
    \widetilde{H}(a)=\sqrt{\widetilde{r}\widetilde{m}^2\widetilde{\phi}_0^2+\Omega_m a^{-3} \over 1-\widetilde{r} a^2  \widetilde{\phi}_0'^2}.
    \label{eq:ha2}
  \end{align}

We leave more details of procedures of solving these background equations to Appendix~\ref{appen:background}.
It is worth noting that according to the definitions in Eqs.~(\ref{def:ndt})~to~(\ref{def:ndm}), there are 2 degrees of freedom for 
the parameters $\widetilde{m}$ and $\widetilde{r}$, to specify the mass and energy scale of the dark energy field $\phi$, respectively. The unknown component in our model, dark energy $\phi$, can be fundamentally characterized by two parameters. One is the shape of its potential $V(\phi)=\msq\phi^2/2$, and the other is the initial value
in our universe, while the properties of the other component (e.g., matter) are considered as known under the standard cosmological model.

In order to fix the dark energy density today, we have the constraint from the present Hubble rate 
by definitions
\begin{align}
   a(\tilde{t}_0)= a(H_0t_0) &\equiv 1,
  \\
   H(\tilde{t}_0) = H(H_0{t_0}) &\equiv H_0,
\end{align}
where $t_0$ is the proper cosmic time for the present epoch.
Inserting this into Eq.~(\ref{eq:tnd1}) actually gives
\begin{align}
  1-\Omega_m= \widetilde{r} \widetilde{m}^2 \left(\widetilde{\phi}_0\Big|_{\tilde{t}
  =\tilde{t}_0}\right)^2+\widetilde{r}\left(\frac{\diff \widetilde{\phi}_0}{\diff \tilde{t}}\bigg|_{\tilde{t}=\tilde{t}_0}\right)^2.
  \label{eq:constr}
\end{align}
Equation~(\ref{eq:constr}) is the necessary condition for specifying the dark energy density observed today when solving the background equations.
Together with Eqs.~(\ref{eq:tnd1}) and (\ref{eq:tnd2}), the system is now prepared for numerical evaluation to obtain the evolution of $a(\tilde{t})$ and $\widetilde{\phi}_0(\tilde{t})$. 
As we are mainly interested in the late-time evolution here, 
we can determine the initial value for independent variables $\tilde{t}$ or $a$ (to be discussed later) manually as a typical value; for example, $a_i = a_d \approx 1/1100$ at the photon decoupling off the last scattering,
by use of Eq.~(\ref{eq:ini_a}). These solutions determine the background evolution that we rely on to solve  the perturbation equations.

It is worth mentioning that Eq.~(\ref{eq:constr}) also provides a particular baseline for choosing the parameters $\widetilde{m}$ and $\widetilde{r}$ from the various parameter spaces,
and that the case for the choice of parameters approximating the $\Lambda$CDM model is
\begin{align}
  \widetilde{r} \widetilde{m}^2 \simeq 1-\Omega_m,
  \label{eq:lcdmparameter}
\end{align}
concerning which more details can be found in Appendix~\ref{appen:background} [see also Eq.~(\ref{eq:phi0-discuss2})].
However, this condition for parameter choice is not mandatory to solve for the system.

\begin{figure}[h]
   \begin{minipage}{0.45\hsize}
     \begin{center}
       \includegraphics[width=0.9\linewidth]{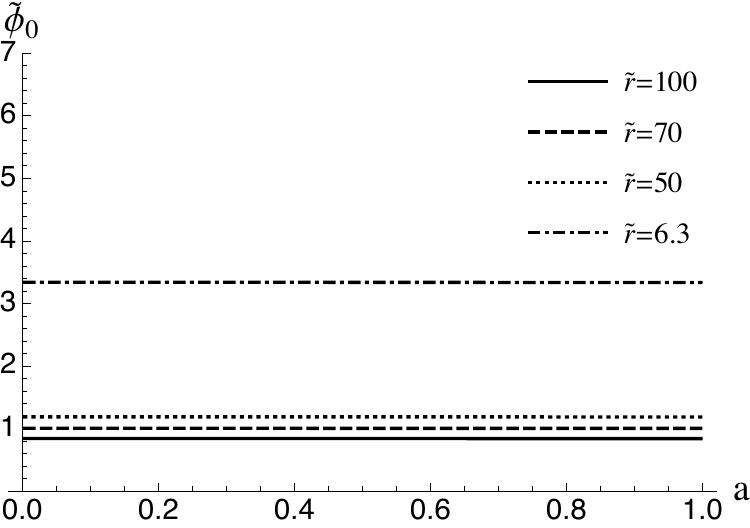}
     \end{center}
         \vspace{-0.cm}
   \end{minipage}
   \begin{minipage}{0.45\hsize}
     \begin{center}
       \includegraphics[width=0.9\linewidth]{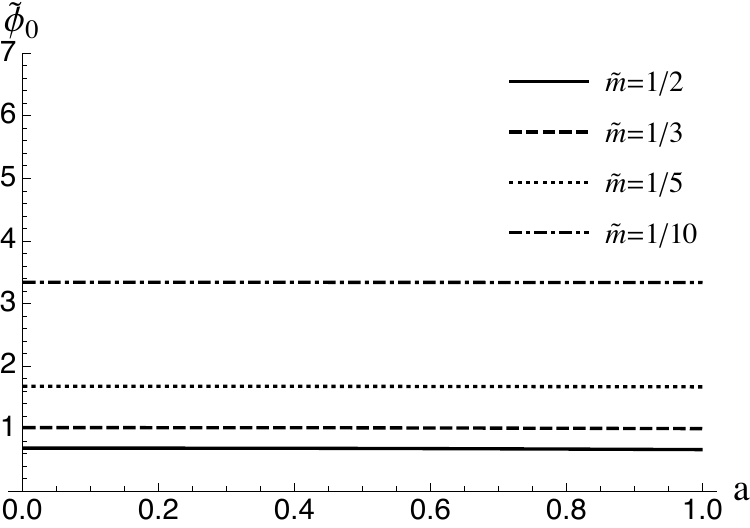}
    \end{center}
         \vspace{-0.cm}      
   \end{minipage}
   \caption{
   Although the parameters in Fig.~\ref{fig:phi0LCDM} seem to be the most natural choices for $\widetilde{m}$ and $\widetilde{r}$, there could be other possibilities. The left panel of this figure shows the impact of the value of $\widetilde{r}$ on the evolution of the background solution $\widetilde{\phi}_0(a)$ as a function of $a$ with fixed $\widetilde{m}=1/10$. 
  According to Eq.~(\ref{def:ndr}), $\widetilde{r}$ reflects the value of the scalar field $\phi$. 
  Hence, this panel demonstrates that the background solution is almost constant, whereas its value, taking $\phi_0 \simgt M_{\rm pl}$, depends on $\widetilde r$. In contrast, the right panel shows the impact of the value of $\widetilde{m}$ on the evolution of the background solution $\widetilde{\phi}_0(a)$ as a function of $a$, where we fixed $\widetilde r=6.3$.
    Recalling Eq.~(\ref{def:ndm}), $\widetilde{m}$ is 
    the parameter of the field mass. These behaviors can be comprehended by Eqs.~(\ref{eq:phi0-discuss1}) and (\ref{eq:phi0-discuss4}), where $\widetilde r$ and $\widetilde m$ act similarly to some rescaling factors of $\widetilde{\phi}_0$. The two parameters correspond to the 2 degrees of freedom for the potential shape of $\phi$, whose parameter space is constrained by the observed dark energy density [see Eq.~(\ref{eq:constr})] and determines the late-time dynamics of $\phi$ until the present epoch, which is assumed to be mild.
  }
  \label{fig:phi0r}
\end{figure}

\begin{figure}[t]
      \includegraphics[width=0.75\linewidth]{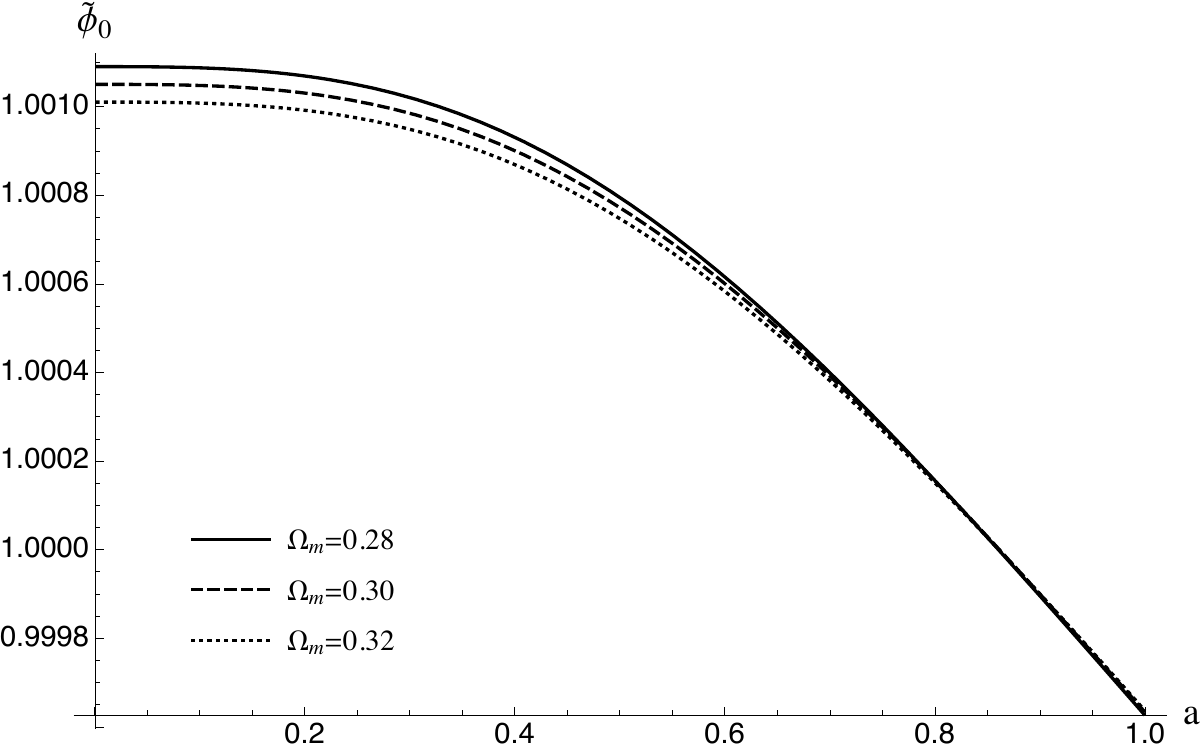}
   \caption{Impact of $\Omega_m$ on the evolution of the background solution $\widetilde{\phi}_0(a)$. Here, we fixed $\widetilde r=70$ and $\widetilde m=1/10$. As expected,  $\Omega_m$ only 
  alter the evolution to a slight extent, suggesting that our model solutions are robust against changes in $\Omega_m$. 
   }
  \label{fig:phi0omegam}
\end{figure}

\begin{figure}[t]
      \includegraphics[width=0.7\linewidth]{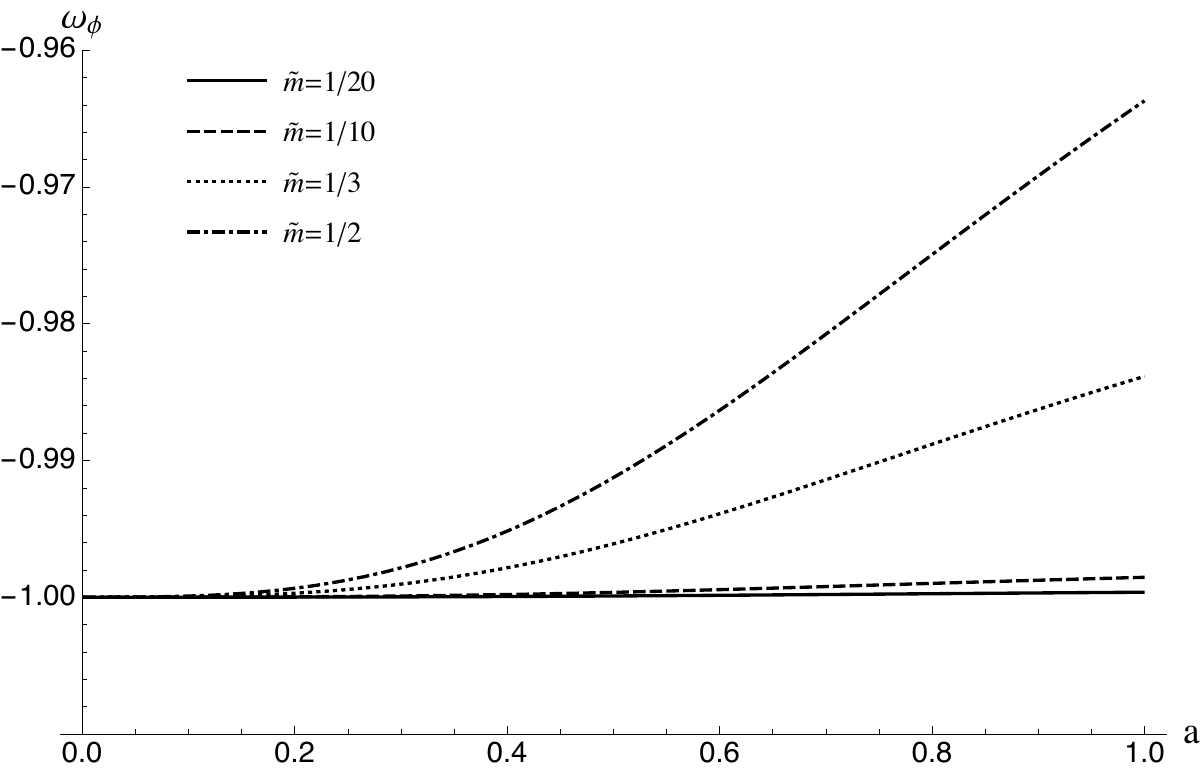}
  \caption{Evolution of the dark energy EoS $\omega_\phi(a)$ with the different sets of the 
  parameters chosen in Fig.~\ref{fig:phi0LCDM}. 
  From Eq.~(\ref{eq:constr}) and Eq.~(\ref{eq:eosparam2}), it is straightforward to see that $\widetilde{r}$ does not affect the EoS of $\widetilde{\phi}_0$. The figure shows the influence of $\widetilde{m}$ on the EoS of $\widetilde{\phi}_0$ with fixed $\Omega_m=0.3$.
   }
  \label{fig:wLCDM}
  \end{figure}

\begin{figure}
    \begin{center}
      \includegraphics[width=0.7\linewidth]{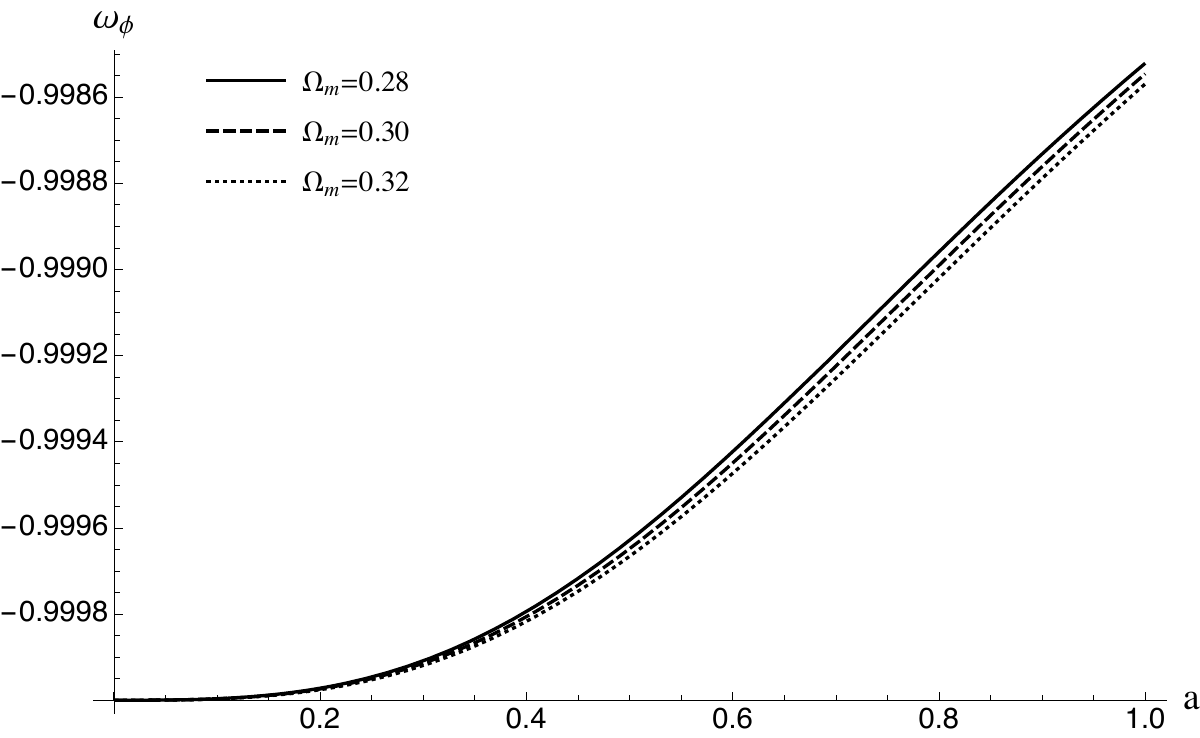}
    \end{center}
  \caption{This figure demonstrates the weak dependence of the EoS on $\Omega_m$ with $\widetilde{m}=1/10$ and $\widetilde{r}=70$ fixed. 
   }
  \label{fig:womegam}
\end{figure}

\begin{figure}[t]
    \begin{center}
      \includegraphics[width=0.7\linewidth]{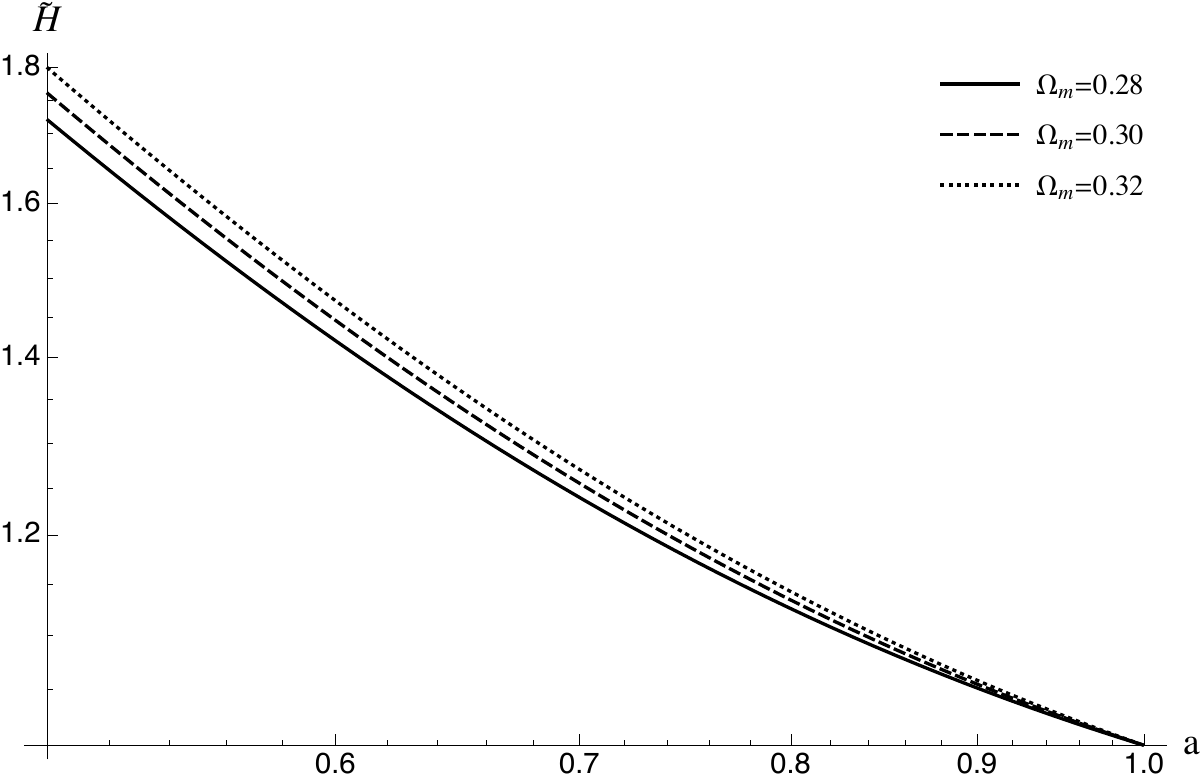}
    \end{center}
    \caption{Examples of the 
    late-time expansion history $\widetilde{H}(a)$ as a function of $a$ for $0.5\leq a\leq 1$ with different sets of $\Omega_m$, 
    $\widetilde{r}\approx 70$  but with $\widetilde{m}=1/10$ fixed.
    For the late-time expansion history, only $\Omega_m$ is important. 
    }
    \label{fig:Homegam}
    \begin{center}
      \includegraphics[width=0.7\linewidth]{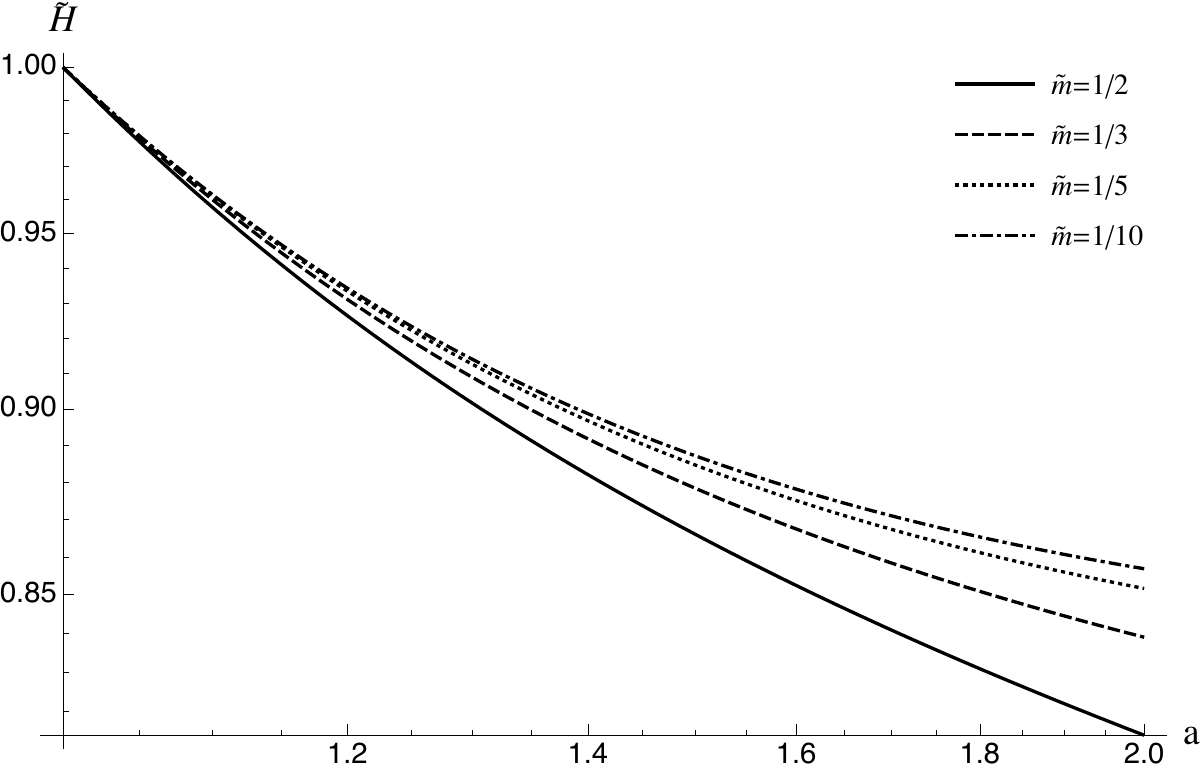}
    \end{center}
    \caption{
    Future evolution of the expansion rate.  
    The parameter $\tilde{m}$ is only important for future expansion when $1\ll a$. The figure demonstrates examples of how the future expansion rate depends on 
    $\widetilde{m}$ for $1<a<2$, where $\widetilde{r}=6.3$ and $\Omega_m=0.3$ are fixed.
    }
    \label{fig:Hmfuture}
\end{figure}

We can now solve for $\widetilde{\phi}_0(a)$ numerically under 2 degrees of freedom for the choice of 
parameters $\widetilde{m}$ and $\widetilde{r}$. Examples of the solutions under the conditions that allow the recovery of the models close to the $\Lambda$CDM universe are presented in Figs.~\ref{fig:phi0LCDM} and \ref{fig:phi0r}.
To investigate the impact of parameter choices on the background solutions more specifically, we also chose other sets of parameters. 
Table~\ref{tab:para} provides the parameter sets adopted in the present paper. The cosmological parameter $\Omega_m$ is related to the fixing of dark energy density at the present epoch, hence slightly affecting $\widetilde{\phi}_0$ if it is not fixed, which is shown in Fig.~\ref{fig:phi0omegam}.
We present some typical figures showing how parameters can affect the equation of state $\omega_\phi$ as a function of $a$ in Figs.~\ref{fig:wLCDM} and \ref{fig:womegam}, where $\widetilde{r}$ is not important (see discussions on $\omega_\phi$ later). Figures~\ref{fig:Homegam} and \ref{fig:Hmfuture} demonstrate the background expansion rate predicted under different parameters. 

We now discuss the behaviors of the background solutions under different parameters. Figure~\ref{fig:phi0LCDM} shows the impact of the 
parameter choice on the behavior of the solution for $\widetilde{\phi}_0(a)$ in the cases following Eq.~(\ref{eq:lcdmparameter}), where models close to the $\Lambda$CDM cosmologies are expected. 
Figure~\ref{fig:phi0r} shows how the parameters $\widetilde{r}$ and $\widetilde{m}$ affect the behaviors of $\widetilde{\phi}_0$, while Fig.~\ref{fig:phi0omegam} shows how $\Omega_m$ can affect $\widetilde{\phi}_0$.
The behaviors of the $\widetilde{\phi}_0$ curves in these figures can be understood as follows.
 From Eqs.~(\ref{eq:tnd1}) and (\ref{eq:constr}) we can see that the parameter $\widetilde{r}$ can actually be 
 absorbed into the amplitude of $\widetilde{\phi}_0$ as a rescaling factor, namely 
 \begin{align}
  \widetilde{m}^2(\sqrt{\widetilde{r}}\widetilde{\phi}_0)^2+\left(\frac{\diff(\sqrt{\widetilde{r}}\widetilde{\phi}_0)}{\diff\tilde{t}} \right)^2
  =\left(\frac{1}{a}\frac{\diff a}{\diff\tilde{t}}\right)^2-\Omega_m a^{-3},
  \label{eq:phi0-discuss1}
 \end{align}
 with
 \begin{align}
  1-\Omega_m=\widetilde{m}^2(\sqrt{\widetilde{r}}\widetilde{\phi}_0\Big|_{a=1})^2+(\sqrt{\widetilde{r}}\widetilde{\phi}_0'\Big|_{a=1})^2.
  \label{eq:phi0-discuss2}
 \end{align}
 These two equations facilitate understanding 
 why changing $\widetilde{r}$ with other parameters fixed only alters the value of $\widetilde{\phi}_0$ without causing a nontrivial difference in the characteristic behaviors of the curves in Fig.~\ref{fig:phi0r}.
 Moreover, as we evaluate $\widetilde{\phi}_0$, choosing the condition in Eq.~(\ref{eq:lcdmparameter}) close to the $\Lambda$CDM model as a baseline for the natural choices of the parameters, 
 \begin{align}
  \frac{\diff(\sqrt{\widetilde{r}}\widetilde{\phi}_0)}{\diff\tilde{t}}\ll1 
  \qquad {\rm or} \qquad 
  \sqrt{\widetilde{r}}\widetilde{\phi}_0'\ll 1 
  \label{eq:phi0-discuss3}
 \end{align}
 always holds.
Hence, it follows Eq.~(\ref{eq:phi0-discuss1}) that
\begin{align}
  (\sqrt{\widetilde{r}}\widetilde{m}\widetilde{\phi}_0)^2
  \simeq\left(\frac{1}{a}\frac{\diff a}{\diff\tilde{t}}\right)^2-\Omega_m a^{-3}.
  \label{eq:phi0-discuss4}
\end{align}
Because of similar arguments for $\widetilde{r}$, we understand that, to some extent, $\widetilde{m}$ also works as a rescaling factor for the background $\widetilde{\phi}_0$, which explains the behavior of $\widetilde{\phi}_0$ in Fig.~\ref{fig:phi0r}. At the same time, the appearance of $\Omega_m$ on the right-hand side of Eq.~(\ref{eq:phi0-discuss4}) explains the dependence of the background solution $\widetilde{\phi}_0$ on $\Omega_m$ in Fig.~\ref{fig:phi0omegam}.

Now, let us discuss the parameter dependence of the dark energy EoS $\omega_\phi(a)$, as shown in
Figs.~\ref{fig:wLCDM} and~\ref{fig:womegam}. We may conclude that the background dark energy EoS $\omega_{\phi}(a)$ is almost 
independent of $\widetilde{r}$; in contrast, $\widetilde{m}$ is the main influencing factor. There is also a slight dependence on the cosmological parameter $\Omega_m$, as shown in Fig.~\ref{fig:womegam}.
These behaviors can be understood using Eqs.~(\ref{eq:eoscpl2})---(\ref{eq:eosparam3}) in Appendix~\ref{appen:EOSCPL} as an analogy to the CPL parametrization. Generally,  $\omega_{\phi}(a)\simeq-1+2\left(1-(a\widetilde{m}^2\widetilde{\phi}_0^2)/(\Omega_m\widetilde{\phi}_0'^2)\right)$ holds for almost all  models; hence, $\widetilde{r}$ does not have an impact on $\omega_\phi$ at the background level, while $\widetilde{m}$ and $\Omega_m$ do affect the dark energy EoS $\omega_\phi$ .

Figure~\ref{fig:Homegam} shows a slight dependence on $\Omega_m$ for the expansion rate $\widetilde{H}(a)$ as a function of the scale factor for $0.5<a<1$, 
while Fig.~\ref{fig:Hmfuture} shows a possible impact on the future expansion rate from the mass parameter $\widetilde{m}$.
To explain these behaviors for $\widetilde{H}(a)$, let us consider the analytic approximation of $\widetilde{H}(a)$ starting from Eq.~(\ref{eq:ha2}). 
For models close to the $\Lambda$CDM model, where $\widetilde{\phi}_0\simeq {\rm const}$ and $\widetilde{\phi}_0'\simeq0$ with Eq.~(\ref{eq:lcdmparameter}), reading $\widetilde{r}\widetilde{m}^2 \simeq 1-\Omega_m$ holds, we have 
\begin{align}
    \widetilde{H}(a) \simeq \sqrt{(1-\Omega_m)\widetilde{\phi}_0^2+\Omega_m a^{-3}},
    \label{Hubbleequation}
\end{align}
which is almost the same as the Hubble equation for the standard $\Lambda$CDM parametrization. Hence, it is obvious that $\Omega_m$ is the dominant parameter for the background expansion history when $0<a<1$.


\begin{figure}[t]
\begin{minipage}{0.55\hsize}
  \begin{center}
    \includegraphics[width=\linewidth]{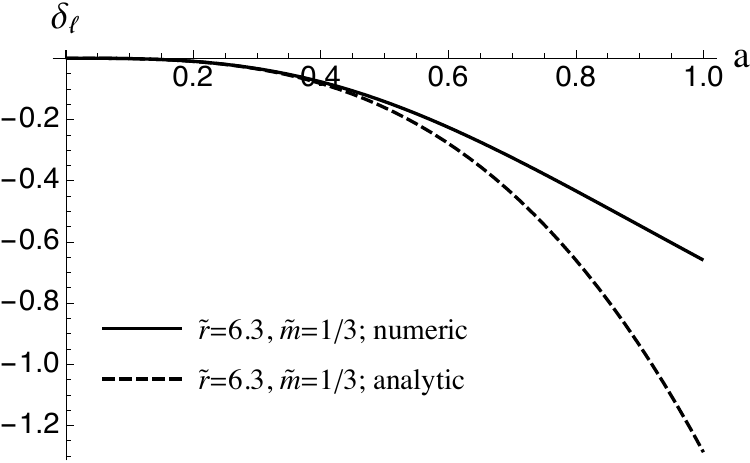}
  \end{center}
      \hspace{.5cm}
\end{minipage}
\begin{minipage}{0.55\hsize}
  \begin{center}
    \includegraphics[width=\linewidth]{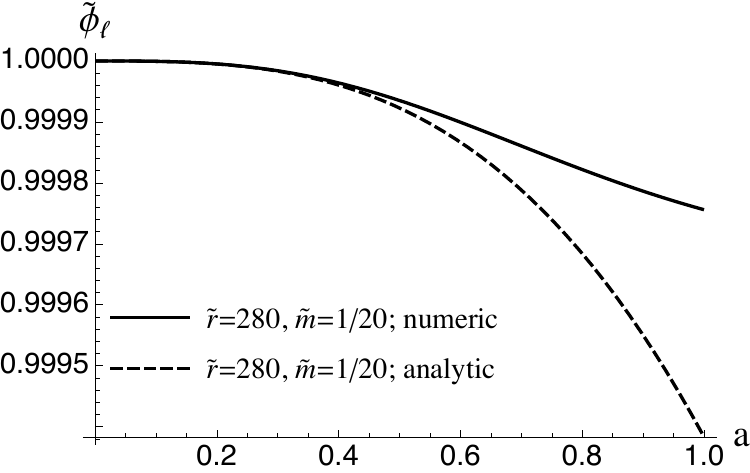}
  \end{center}
      \vspace{-0.cm}      
\end{minipage}
\caption{Comparison of the evolution of $\delta_\ell$ and $\widetilde \phi_\ell$ between the analytic approximation (dashed curve) by Eqs.~(\ref{eq:ini_deltal}) and (\ref{eq:ini_phil}), and the exact numerical solutions (solid curves). Here, we adopt $\widetilde{r}=70$ and $\widetilde{m}=1/10$ for $\delta_\ell$, and 
$\widetilde{r}=280$ and $\widetilde{m}=1/20$ for $\tilde{\phi}_\ell$ as examples. We checked the validity of the analytic approximations for other values of $\widetilde{m}$ and $\widetilde{r}$ adopted in Table I.
The deviation between the analytic approximation and the numerical solution starting around $a\gtrsim 0.5$ arises from the emerging domination of dark energy, which breaks down the analytic approximation obtained from the initial condition of matter domination.
}
\label{fig:numer-analytic} 
\end{figure}


\begin{figure}[t]
      \includegraphics[width=0.7\linewidth]{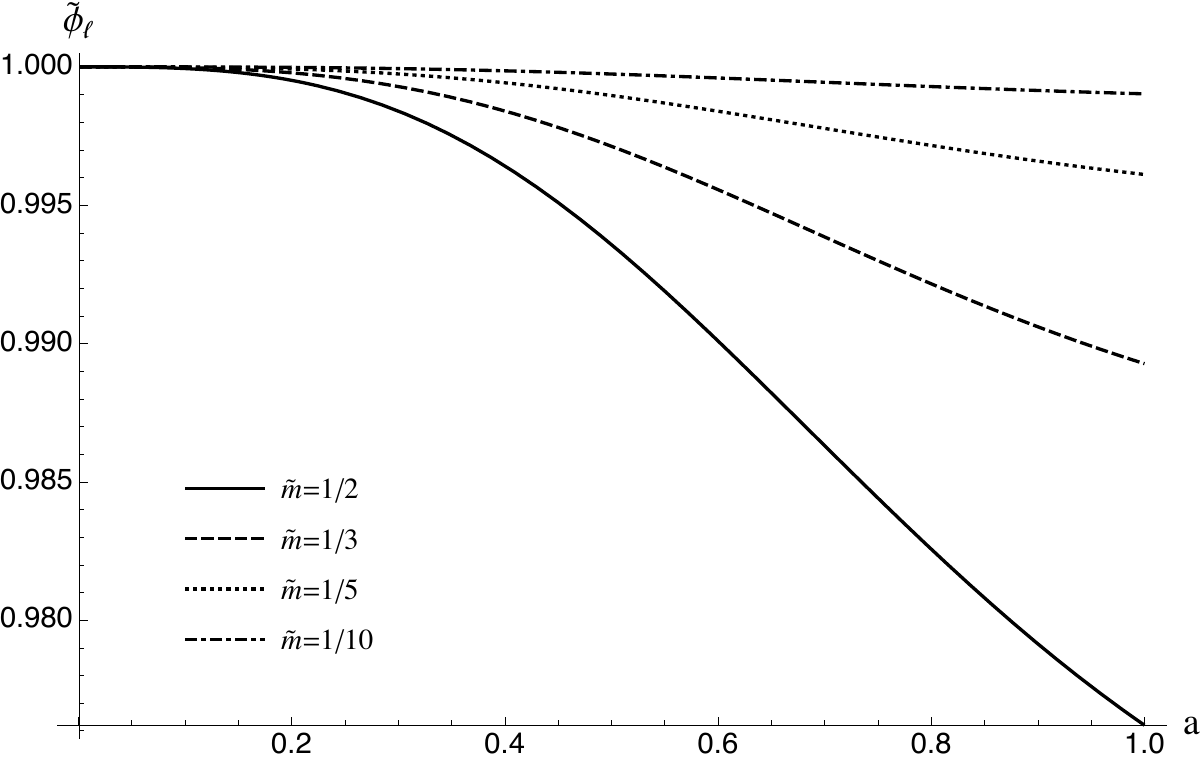}
  \caption{Numerical solutions for the perturbation for $\tilde{\phi}_\ell(a)$, with the different values of parameter $\widetilde{m}$, where $\Omega_m=0.3$ and $\widetilde{r}=6.3$ are fixed. 
  }
  \label{fig:philm}
\end{figure}


\subsection{Equations governing first-order perturbations}
In the previous subsection, we have solved for the background, and on the basis of the background solutions we now consider the numerical solution for the first-order perturbation equations in Eqs.~(\ref{eq:11})--(\ref{eq:16}) that we are interested in.

We define the perturbation to dark matter density as
\begin{align}
  \rho_\ell\equiv\rho_0\delta_\ell, 
\end{align}
together with the following quantity associated with the velocity as
\begin{align}
  \widetilde{V_{\ell}}\equiv H_0 V_{\ell}.
\end{align}

Then, we can utilize the Friedmann equation relation in Eqs.~(\ref{eq:11})---(\ref{eq:16}) to eliminate quantities such as $\rho_0$ and $\rho_\ell$, and 
use $\delta_\ell$ to characterize the first-order matter perturbations as 
\begin{align}
  \rho_0(a)=3H_0^2\Omega_m a^{-3} M_{\rm pl}^2,
  \label{eq:rho0}
\end{align}
Thus the dimensionless differential equations as functions of $\tilde{t}$ will be
\begin{gather}
  {\partial {\delta}_\ell \over \partial \tilde t}+3{\partial{\Phi}_\ell \over \partial \tilde{t}}=0,
  \label{eq:1st1}
  \\
  {\partial^2\widetilde{\phi}_\ell \over \partial \tilde{t}^2}
  + {3 \over a}{\partial a\over \partial \tilde{t}}{\partial \widetilde{\phi}_\ell \over \partial \tilde{t}}+\widetilde{m}^2\widetilde{\phi}_\ell
  -2\Psi_\ell {\partial^2\widetilde{\phi}_0 \over \partial \tilde{t}^2}
  -{6 \Psi_\ell \over a}{\partial a\over \partial \tilde{t}}{\partial \widetilde{\phi}_0 \over \partial \tilde{t}}
  +{\partial \over \partial \tilde{t}}(3\Phi_\ell-\Psi_\ell)
  {\partial \widetilde{\phi}_0 \over \partial \tilde{t}}=0,
  \label{eq:1st3}
  \\
  {\partial\widetilde{V}_\ell \over \partial \tilde{t}}-\Psi_\ell=0,
  \label{eq:1st2}
  \\    
  -{2\over a}{\partial a \over \partial \tilde{t}}\Psi_\ell
  +2{\partial \Phi_\ell \over \partial \tilde{t}}+ 3 \widetilde{V}_\ell \Omega_m a^{-3}
  +6\widetilde{r} \widetilde{\phi}_\ell{\partial\widetilde{\phi}_0 \over \partial\tilde{t}}=0,
  \\
 6 ({1 \over a}{\partial a\over \partial \tilde{t}})^2 \Psi_\ell -6({1 \over a}{\partial a\over \partial \tilde{t}}){\partial{\Psi}_\ell \over \partial \tilde{t}}+3\Omega_m a^{-3}\delta_\ell
  +6\widetilde{r}\left(
    \widetilde{m}^2\widetilde{\phi}_0\widetilde{\phi}_\ell+{\partial\widetilde{\phi}_0 \over \partial\tilde{t}}{\partial\widetilde{\phi}_\ell \over \partial\tilde{t}}-\Psi_\ell ({\partial\widetilde{\phi}_0 \over \partial\tilde{t}})^2
    \right)=0,
  \\
  \left(({1 \over a}{\partial a\over \partial \tilde{t}})^2+{2 \over a}{\partial^2 a\over \partial \tilde{t}^2}\right)\Psi_\ell
  +{1 \over a}{\partial a\over \partial \tilde{t}}{\partial \over \partial \tilde{t}}\left(\Psi_\ell-3\Phi_\ell\right)
  -{\partial^2\Phi_\ell \over \partial \tilde{t}^2}+3\widetilde{r}
  \left(\widetilde{m}^2\widetilde{\phi}_0\widetilde{\phi}_\ell-{\partial\widetilde{\phi}_0 \over \partial\tilde{t}}{\partial\widetilde{\phi}_\ell \over \partial\tilde{t}}+\Psi_\ell({\partial\widetilde{\phi}_0 \over \partial\tilde{t}})^2\right)=0.
  \label{eq:1st6}
\end{gather}
Notice that from Eq.~({\ref{eq:1st1}})
\begin{align}
  \delta_\ell+3{\Phi}_\ell={\rm const},
  \label{eq:inidelta}
\end{align}
where the constant is presumed to be zero as we assume that the superhorizon
perturbations of the scalar field are the isocurvature perturbations. 
Then, we assume the initial values
\begin{align}
\delta_\ell(0)=\Phi_\ell(0)=0.
  \label{eq:inidelta2}
\end{align}
As in the case of the supercurvature mode dark energy \cite{scmde1}, 
if we adopt the general condition that anisotropic stress is negligible, which reads
\begin{align}
  {\Phi}_\ell+{\Psi}_\ell\simeq0,
\end{align} 
we can eliminate ${\Phi}_\ell$ and ${\Psi}_\ell$ using $\delta_\ell$ and $\partial \widetilde{V}_\ell/ \partial \tilde{t}$ using Eq.~(\ref{eq:1st2}).
Finally, we will have two equations for $\delta_\ell$ and $\tilde{\phi}_\ell$ to solve, whose explicit forms are long and trivial 
hence we omit them here. 
We note that our analysis is based on the conformal Newtonian (longitudinal) gauge, 
which is widely used in various analyses of cosmological perturbations. 
It is known that the conformal Newtonian gauge
leaves no residual gauge freedom except for the long wavelength mode of $k = 0$.
The effect of the inhomogeneities of our
dark energy model is the isocurvature perturbations 
in the long wavelength limit. We consider that the gauge freedom is 
fixed for the dipole and quadrupole modes with nonzero small $k$; however, the possibility of 
contamination by the gauge modes with $k=0$ could be mentioned. 

Again, we need to consider the initial conditions for which we solve the equations in the limit
$a\ll1$ in an analytic manner.
First, recalling the definition of Eqs.~(\ref{def:field})~and~(\ref{def:ndphi}), we can generalize the dimensionless quantities as:
\begin{align}
  \phi
  &\equiv
  \bar{\phi}_0(\widetilde{\phi}_0+\epsilon_1\widetilde{\phi}_1\sum_m P_i^{(m)}x^{i}+\epsilon_2\widetilde{\phi}_2\sum_m P^{(m)}_{ij} x^i x^j)
\end{align}

In the limit $a\ll 1$ ($t \rightarrow 0, \widetilde t \rightarrow 0$),
we may assume the power law form for the perturbations
\begin{align}
  \delta_\ell &\equiv A_1\tilde{t}^{\alpha},
  \label{assump:delta}
  \\
  \widetilde{\phi}_\ell &\equiv \D+D_1\tilde{t}^{\gamma}.
  \label{assump:phil}
\end{align}
Furthermore, Eq.~(\ref{eq:iniphi}) gives
\begin{align}
  \widetilde{\phi}_0(\tilde{t})= C_1 \frac{\sin(\widetilde{m}\tilde{t})}{\widetilde{m}\tilde{t}}\approx
  C_1 (1-\frac{\widetilde{m}^2\tilde{t}^2}{6})
  \equiv  F(1-\frac{\widetilde{m}^2\tilde{t}^2}{6}).
  \label{assump:phi}
\end{align}
For a given $\widetilde{m}$ and $\widetilde{r}$, we solve for the background and fix the value for $C_1$ or $F$ in Sec.~\ref{sec:numer_background}. 
We may take $F$ as a known quantity here.
For scale factor $a$, recall that Eq.~(\ref{eq:ini_a}) is the background analytical approximation as 
\begin{gather}
  a= \left(\frac{9}{4}\Omega_m \right)^{\frac{1}{3}} \tilde{t}^{\frac{2}{3}}\equiv B \tilde{t}^{\frac{2}{3}} ,
~~~  \qquad \tilde{t} = \left({\frac{a}{B}}\right)^{3 \over 2}.
  \nonumber
\end{gather}

Inserting the ansatz Eqs.~(\ref{assump:delta})---(\ref{assump:phi}) into Eqs.~(\ref{eq:1st3})---(\ref{eq:1st6})
will give us equations as a function of $a$ or $\tilde{t}$ relating the unknown coefficients $\alpha$,$\gamma$,$A_1$,$\D$, $D_1$ that we want to explore.
For the limit $a \to 0$ or $\tilde{t} \to 0$, by looking at the leading order of $a$ for each equation, we have

\begin{gather}
  \alpha=\gamma=2,
  \\
  D_1= -{1\over 6}\widetilde{m}^2 \D,
  \label{coefd1}
  \\
  A_1= - {27\over 22}\widetilde{m}^2 \widetilde{r} F \D,
  \label{coefa1}
\end{gather}
where $\D$ may be understood as the amplitude of each mode of the perturbations as $\epsilon_1$ and $\epsilon_2$, which will be constrained later with the observational data. 
For now, $\D=1$ may be set for the numerical solution. 

Further, the analytic approximations for the evolution of the perturbations in the limit $a\ll 1$ ($t \rightarrow 0, \widetilde t \rightarrow 0$) are found as
\begin{align}
  \delta_\ell &\simeq - {27\over 22}\D\widetilde{m}^2 \widetilde{r} F \tilde{t}^{2}= - {27\over 22}\widetilde{m}^2 \widetilde{r} F \tilde{t}^{2},
  \label{eq:ini_deltal}
  \\
  \widetilde{\phi}_\ell &\simeq \D \left(1 -{1\over 6}\widetilde{m}^2\tilde{t}^{2}\right)= 1 -{1\over 6}\widetilde{m}^2\tilde{t}^{2},
  \label{eq:ini_phil}
\end{align}
allowing us to set the proper initial conditions for $\delta_\ell$ and $\widetilde{\phi}_\ell$. 
The equations using $a$ and $\tilde{t}$ as an independent variables are mutually transformable using Eq.~(\ref{eq:ini_a}),
as was done in Sec.~\ref{sec:numer_background}. 
The analytical solution of the first-order equations in Eqs.~(\ref{eq:11})--(\ref{eq:16}) for the other quantities can be
found in a similar way as
\begin{align}
    &\Phi_\ell\simeq-\Psi_\ell\simeq+\frac{9}{22}\D\widetilde{m}^2 \widetilde{r} F \tilde{t}^{2}=+\frac{9}{22}\widetilde{m}^2 \widetilde{r} F \tilde{t}^{2} ,
    \label{eq:ini_PsiPhi}
    \\
    &\widetilde{V}_\ell \simeq -\frac{3}{22}\D\widetilde{m}^2 \widetilde{r} F \tilde{t}^{3}= -\frac{3}{22}\widetilde{m}^2 \widetilde{r} F \tilde{t}^{3}.
    \label{eq:ini_v}
\end{align}
We notice that $\delta_\ell$ and $\Psi_\ell$ are negative values, which correspond to the positive values of $\widetilde{\phi}_\ell$ in Eq.~(\ref{eq:ini_phil}). Physically, this means that an increase in dark energy $\phi$
makes the matter density perturbations $\delta_\ell$ (curvature potentials $\Phi_\ell$) negative (positive). 

The first-order equations (\ref{eq:11})---(\ref{eq:16}) can be
solved in an exact manner using a numerical method. We present examples of the numerical solutions for perturbations $\tilde{\phi}_\ell(a)$ and $\delta_\ell(a)$ in Fig.~\ref{fig:numer-analytic}, where 
we adopted $\D=1$ with the same typical parameter sets $\widetilde{r}$ and $\widetilde{m}$ chosen in Sec.~\ref{sec:numer_background}. The consistency between the analytic approximations in Eqs.~(\ref{eq:ini_deltal}) and (\ref{eq:ini_phil}) (dashed line) with the numerical results (solid line) when $a\lesssim0.5$ corresponding to the matter-dominant initial condition is also demonstrated 
in Fig.~\ref{fig:numer-analytic}, while the analytic approximation deviates from the numerical solution when $a\gtrsim0.5$.

We show how $\widetilde{m}$ affects the solution $\tilde{\phi}_\ell$ in Fig.~\ref{fig:philm}. 
It should be noted that there is a slight dependence on $\Omega_m$ for $\tilde{\phi}_\ell$, similar to the behavior of $\widetilde{\phi}_0$ in Fig.~\ref{fig:phi0omegam}. 
The behaviors of $\tilde{\phi}_\ell$ can be roughly understood from Eq.~(\ref{eq:ini_phil}), which is valid for $a \lesssim 0.5$. 
Here $\widetilde{m}$ is important for the evolution of $\tilde{\phi}_\ell$, 
whereas $\widetilde{r}$ is not. On the other hand, Eq.~(\ref{eq:12}) indicates that the solution of $\tilde{\phi}_\ell$ depends on $\widetilde{\phi}_0$; hence, it slightly depends on $\Omega_m$, which can be understood by a discussion similar to that on the behavior of $\widetilde{\phi}_0$ in Sec.~\ref{sec:numer_background} [see Eq.~(\ref{eq:phi0-discuss4})].

\begin{figure}
\begin{minipage}{0.45\hsize}
  \begin{center}
    \includegraphics[width=\linewidth]{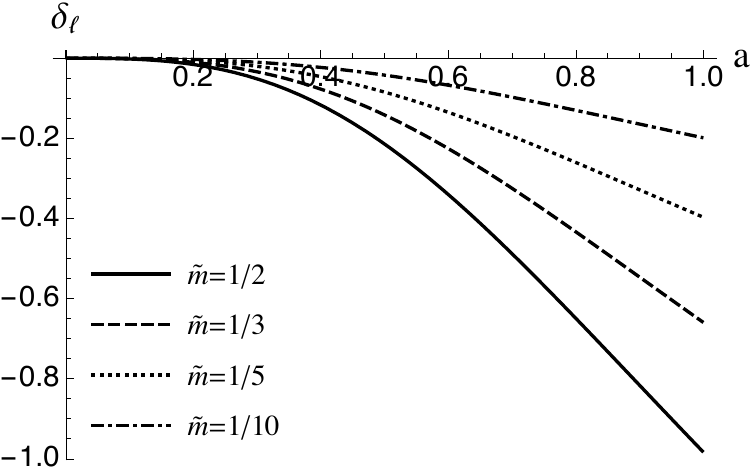}
  \end{center}
      \vspace{-0.cm}
\end{minipage}
\begin{minipage}{0.45\hsize}
  \begin{center}
    \includegraphics[width=\linewidth]{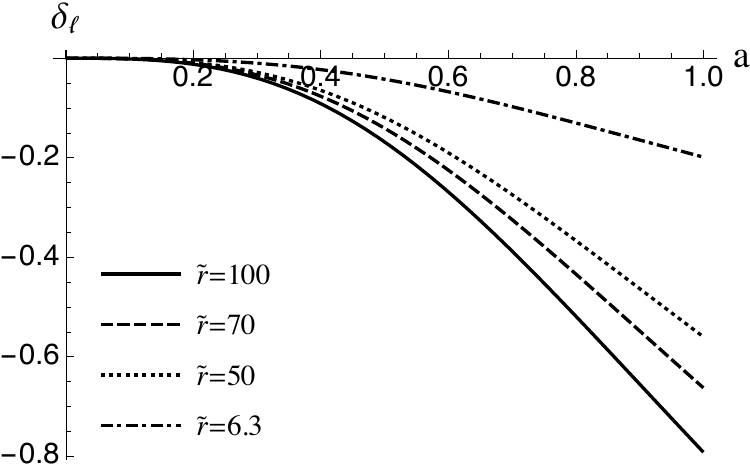}
  \end{center}
      \vspace{-0.cm}      
\end{minipage}
\begin{minipage}{0.45\hsize}
  \begin{center}
    \includegraphics[width=\linewidth]{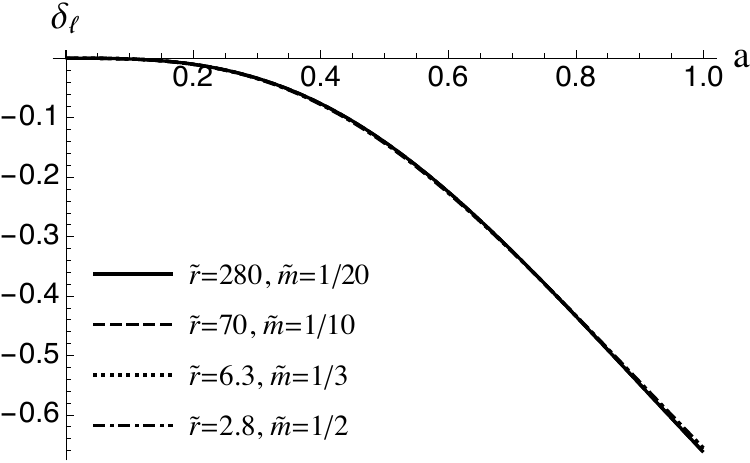}
  \end{center}
      \vspace{-0.cm}
\end{minipage}
\begin{minipage}{0.45\hsize}
  \begin{center}
    \includegraphics[width=\linewidth]{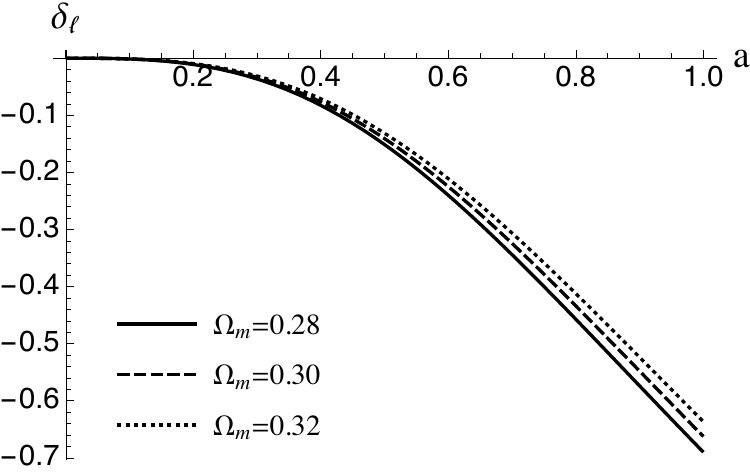}
  \end{center}
      \vspace{-0.cm}      
\end{minipage}
\caption{
Numerical solutions for the matter perturbation $\delta_\ell$. 
The upper left and upper right panels demonstrate the dependence of $\delta_\ell$ on 
$\widetilde{m}$ and $\widetilde{r}$, respectively. 
The lower left panel assumes the same value of $\Omega_m=0.3$, 
while the lower right panel assumes slightly different values of 
$\Omega_m$, 
where $\widetilde{r}=70$ and $\widetilde{m}=1/10$ are fixed. 
The lower panels show that $\delta_\ell$ will 
be almost independent of $\widetilde{r}$ or $\widetilde{m}$ values, 
as long as they satisfy Eq.~(\ref{eq:lcdmparameter}).
}
  \label{fig:delta-lcdm-omegam}
\end{figure}

The dependence on the parameters for $\delta_\ell$ is shown in Fig.~\ref{fig:delta-lcdm-omegam}. From Eq.~(\ref{eq:ini_deltal}), we can conclude that $\widetilde{m}$ and $\widetilde{r}$ affect $\delta_\ell$, which is demonstrated in the upper left panel and the upper right panel of Fig.~\ref{fig:delta-lcdm-omegam}, respectively. However, for natural choices mimicking the standard $\Lambda$CDM scenario, satisfying Eq.~(\ref{eq:lcdmparameter}), the coefficient $F\approx1$ holds; hence, we have $\delta_\ell \simeq -(27/22)(1-\Omega_m)\tilde{t}^2$, which explains the behavior of $\delta_\ell$ in the lower panels of Fig.~\ref{fig:delta-lcdm-omegam}.

\section{Applications}
\label{sec:appli}
In this section, we consider two applications of our model for CMB temperature fluctuations and luminosity distance. 
The first is the integrated Sachs-Wolfe (ISW) effect~\cite{scmde1,WHuthesis}.  Some aspects of this effect were 
investigated in a previous paper \cite{scmde1}, which relies on 
the statistical argument based on the two-point correlation function.
We revisit this problem by applying the formulations developed in the present study. The second is the impact on the luminosity distances, which is related to the observations of SNe Ia.  

As noted following the definition of $\phi$, Eq.~(\ref{def:field}), $\epsilon_\ell$
was introduced to explicitly express the order of perturbations that are small, and was related to the coordinates we choose to define the multipoles of the perturbations in Eqs.~(\ref{def:Psi})--(\ref{def:velo}). These amplitudes of perturbations will be taken unity, that is, $\epsilon_\ell \sim \D_{(\ell m)} \equiv 1$ (see also Eqs.~(\ref{assump:delta}), (\ref{assump:phil}), (\ref{coefd1}) and (\ref{coefa1})), for the purpose of numerical evaluation, where most importantly we are interested in the evolution of the perturbations. These amplitudes of the perturbations will be constrained by reintroducing other
parameters $\varepsilon_1$ and $\varepsilon_2$ when comparing with the actual CMB multipoles observed.

\subsection{CMB temperature fluctuations}
\label{sec:appli_CMB}
Through the ISW effect, the perturbations to the metric caused by 
the large-scale inhomogeneities of the dark energy $\phi$ affect the observations of the CMB anisotropies.
By using the relation between the comoving distance and the conformal time on the photon's path on the background, $\chi=\eta_0-\eta$, 
 we can evaluate the ISW effect on the temperature fluctuations of the CMB as
\begin{align}
  {\Delta T\over T}
  \simeq &
  2\int_{\eta_d}^{\eta_0}\diff\eta\left({\partial\Psi(\eta,\chi,\theta,\varphi)\over\partial\eta}\right)\Bigg|_{\chi=\eta_0-\eta}
  \nonumber
  \\
  =& 
  2 \int_{\eta_d}^{\eta_0} \diff\eta \left(
  \sum^3_{m=1}{\partial\Psi_{1(m)}(\eta)\over\partial\eta} P^{(m)}_i x^i + 
  \sum^5_{m=1}{\partial \Psi_{2(m)}(\eta)\over\partial\eta} P^{(m)}_{ij} x^i x^j \right) \Bigg|_{\chi=\eta_0-\eta}
  \nonumber
  \\
  =& 
  2 
  \int_{\eta_d}^{\eta_0} \diff\eta  \left({\partial\Psi_{1(m)}(\eta) \over\partial\eta}  \sum^3_{m=1} P^{(m)}_i x^i \right)\Bigg|_{\chi=\eta_0-\eta} 
  +2
  \int_{\eta_d}^{\eta_0}  \diff\eta  \left({\partial\Psi_{2(m)}(\eta)\over\partial\eta}  \sum^5_{m=1} P^{(m)}_{ij} x^i x^j \right)\Bigg|_{\chi=\eta_0-\eta},
  \label{eq:tempfluc1}
\end{align}
  where $\eta_d$ denotes the era of the photon decoupling.
  In the last line of Eq.~(\ref{eq:tempfluc1}), we used the Einstein summation convention
  with respect to the index of $m$.
We note that $\Psi_{\ell(m)}$, which are denoted as $\Psi_{\ell}$ with the index $m$ omitted in the previous section for simplicity, are only functions of 
the conformal time $\eta$. 
It can also be confirmed that the matrices $P_{ij}^{m}$ and $P_{i}^{m}$ introduced in Sec.~\ref{sec:basic} are related to the 
real basis spherical harmonics $Y^{m}_\ell(\theta,\varphi)$ (see Appendix~\ref{appen:matrix}).
By utilizing the relation in Eqs.~(\ref{def:y1m}) and (\ref{def:y2m}), it follows that
\begin{align}
  {\Delta T\over T}
  &=
 2\sum_m
  \int_{\eta_d}^{\eta_0} \diff\eta  \left({\partial\Psi_{1(m)} \over\partial\eta} \chi  Y_{\ell=1}^{(m)}(\theta,\varphi) \right)\Bigg|_{\chi=\eta_0-\eta} 
 +
 2\sum_m\int_{\eta_d}^{\eta_0} \diff\eta  \left({\partial\Psi_{2(m)} \over\partial\eta} \chi^2 Y_{\ell=2}^{(m)}(\theta,\varphi) \right)\Bigg|_{\chi=\eta_0-\eta} 
 \nonumber
 \\
 &\equiv
 2\sum_{\ell=1}^2 \sum_{m=1}^{2\ell+1}
 Q_{\ell(m)} Y_{\ell}^{(m)}(\theta,\varphi),
 \label{eq:tempfluc2}
\end{align}
 with 
\begin{align}
 &Q_{\ell (m)}\equiv \int_{\eta_d}^{\eta_0} \diff\eta (\eta_0-\eta)^{\ell} {\partial\Psi_{\ell(m)} \over\partial\eta}.
\end{align}
defined.
Because we have obtained the evolution of the perturbation $\Psi_{\ell(m)}$ in the previous numerical solution in Sec.~\ref{sec:numer}, $Q_{\ell(m)}$ can be numerically evaluated.

On the other hand, the angular two-point correlation function can be written in multipole expansion as~\cite{Bielewicz2004}
\begin{align}
  \langle {\Delta T\over T}(\bm \gamma) {\Delta T\over T } (\bm \gamma') \rangle 
=\sum_\ell \frac{2\ell+1}{4\pi}C_\ell P_\ell(\cos\theta),
\end{align}
where $\bm \gamma$ and $\bm \gamma'$ represent different unit line-of-sight directions with included angle $\theta$, i.e., $\bm \gamma \cdot \bm\gamma'=\cos\theta$.
The angular power spectrum $C_\ell$ is defined by the ensemble of squared expansion coefficients as follows:
\begin{gather}
  C_\ell \equiv \displaystyle{\sum_{m=1}^{2\ell+1} |A_{\ell m}|^2\over 2\ell+1},
\end{gather}
where the coefficients are defined by
\begin{gather}
  {\Delta T\over T}=\sum_\ell \sum_{m=1}^{2\ell+1} A_{\ell m} Y_{\ell}^{(m)}(\theta,\varphi).
  \label{def:tempfluc1}
\end{gather}
Here we used $1\leq m\leq 2\ell+1$ to denote the magnetic quantum number.
By comparing Eq.~(\ref{eq:tempfluc2}) with (\ref{def:tempfluc1}), we find that
\begin{align}
 A_{\ell m}=2 
 Q_{\ell(m)}= 
 2\left(\int_{\eta_d}^{\eta_0} \diff\eta (\eta_0-\eta)^\ell {\partial\Psi_{\ell(m)} \over\partial\eta} \right).
\end{align}

A constraint on our model from the observational CMB power spectrum is $
C_\ell \leq C_\ell^{\rm obs}$, 
which means that the contribution of the large-scale mode perturbations to the CMB power spectrum multipoles should not exceed what is actually observed, because 
there may be other sources contributing to the anisotropies, as long as cancellations do not occur. 
Consequently, we have two constraints from the $\ell=1$ dipole and the $\ell=2$ quadrupole, respectively, as
\begin{align}
  {4\sum_{m=1}^{2\ell+1} Q_{\ell(m)}^2\over 2\ell+1} \leq C_\ell^{\rm obs}.
\end{align} 
Thanks to the Planck Legacy Archive,~\footnote{Based on observations obtained with Planck (http://www.esa.int/Planck), an ESA science mission with instruments and contributions directly funded by ESA Member States, NASA, and Canada.}
we can apply the upper limit of the observational data as $C_1^{\rm obs}<6.3\times10^{-6}$ and $C_2^{\rm obs}<(2\pi/6)\times(1.0\times10^{-10})$ to put 
constraints on the amplitudes of the perturbations.

For example, for both parameter sets $(\widetilde{r}=70, \widetilde{m}=1/10)$ and $(\widetilde{r}=6.3, \widetilde{m}=1/3)$, or, more generally, for models close to $\Lambda$CDM sets labeled with Nos.~(1),~(2),~(7),~(8) in Table~\ref{tab:para}, where the condition in Eq.~(\ref{eq:lcdmparameter}) is satisfied, the calculations on $Q_{1(m)}$ and $Q_{2(m)}$ give consistent results as 
\begin{align}
  &Q_{1(m)}=-1.1\times10^{-1}\D_{(1 m)},
  \\
  &Q_{2(m)}=-9.0\times10^{-2}\D_{(2 m)},
\end{align}
where the amplitude of the perturbations for each mode 
$\D_{(\ell m)}$ is recovered,  
which lead to the following constraints 
\begin{align}
  &\varepsilon_1\equiv
  \left[\sum_{m=1}^{2\ell+1} \D_{(1 m)}^2\over 2\ell+1\right]^{1/2}
  \leq 1.2\times10^{-2},
  \label{constr1}
  \\
  &\varepsilon_2\equiv\left[\sum_{m=1}^{2\ell+1} \D_{(2 m)}^2\over 2\ell+1\right]^{1/2}
  \leq 5.7\times10^{-5} ,
  \label{constr2}
\end{align}
because both parameter sets mimic the cosmology close to a $\Lambda$CDM model to yield the observational constraints safely.
We also present numerical evaluations with different parameter choices in Table~\ref{tab:para}.

\begin{table}[h]
\begin{center}
\begin{tabular}{c||cc||cc|cc|cc||c}
\hline
\hline
$\rm{No.}$&$(\widetilde{r},\widetilde{m})$ & $\Omega_m$ & $Q_{1(m)}$ &  $Q_{2(m)}$ & $\varepsilon_1^{\rm max}$ & $\varepsilon_2^{\rm max}$ & $F_{S1(m)}(z=3)$ & $F_{S2(m)}(z=3)$ & $H_0\eta_0$\\
\hline
$(1)$&$(70, 1/10)$ & 0.30 & -0.107 & -0.0895 & $1.17\times10^{-2}$ & $5.72\times10^{-5}$ & -0.0462 & -0.0693  & 3.19\\
$(2)$&$(6.3, 1/3)$ & 0.30 & -0.107 & -0.0896 & $1.17\times10^{-2}$ & $5.71\times10^{-5}$ & -0.0462 & -0.0692 & 3.19\\
$(3)$&$(50, 1/10)$ & 0.30 & -0.0904 & -0.0757 & $1.39\times10^{-2}$ & $6.76\times10^{-5}$ & -0.0390 &-0.0586 & 3.19\\
$(4)$&$(100, 1/10)$ & 0.30 & -0.128 & -0.107 & $9.82\times10^{-3}$ & $4.78\times10^{-5}$ & -0.0552 & -0.0828 & 3.19\\
$(5)$&$(6.3, 1/5)$ & 0.30 & -0.0642 & -0.0537 & $1.96\times10^{-2}$ & $9.52\times10^{-5}$ & -0.0277 & -0.0416 & 3.19\\
$(6)$&$(6.3, 1/10)$ & 0.30 & -0.0321 & -0.0269 & $3.91\times10^{-2}$ & $1.91\times10^{-4}$ & -0.0138 & -0.0208 & 3.19\\
$(7)$&$(2.8, 1/2)$ & 0.30 & -0.107 & -0.0897 & $1.18\times10^{-2}$ & $5.70\times10^{-5}$ & -0.0463 & -0.0692 & 3.19\\
$(8)$&$(280, 1/20)$ & 0.30 & -0.107 & -0.0895 & $1.17\times10^{-2}$ & $5.72\times10^{-5}$ & -0.0461 & -0.0693 & 3.19\\
$(9)$&$(72, 1/10)$ & 0.28 & -0.116 & -0.100 & $1.08\times10^{-2}$ & $5.11\times10^{-5}$ & -0.0503 & -0.0770 & 3.28\\
$(10)$&$(68, 1/10)$ & 0.32 & -0.0985 & -0.0803 & $1.27\times10^{-2}$ & $6.37\times10^{-5}$ & -0.0425 & -0.0626 & 3.11\\
$(11)$&$(1/70, 1/10)$ & 0.30 & -0.00153 & -0.00128 & $8.21\times10^{-1}$ & $4.00\times10^{-5}$ & -0.000659 & -0.000990 & 3.19\\
$(12)$&$(6.3, 1/2)$ & 0.30 & -0.160 & -0.135 & $7.83\times10^{-3}$ & $3.80\times10^{-5}$ & -0.0694 & -0.104 & 3.19\\
$(13)$&$(70, 1/10)$ & 0.32 & -0.100 & -0.0815 & $1.25\times10^{-2}$ & $6.28\times10^{-5}$ & -0.0431 & -0.0635 & 3.11\\
$(14)$&$(70, 1/10)$ & 0.28 & -0.115 & -0.0988 & $1.09\times10^{-2}$ & $5.18\times10^{-5}$ & -0.0496 & -0.0759 & 3.28\\
\hline
\hline
\end{tabular}
\caption{
Numerical results with different model parameters $(\widetilde{r},\widetilde{m})$ and cosmological parameter $\Omega_m$.
The models close to the $\Lambda$CDM model are labeled as Nos.~(1),~(2),~(7),~(8),~(9),~(10),~(13), and (14). 
Within these models, Nos.~(1),~(2),~(7),~(8),~(9), and (10) satisfy the condition in Eq.~(\ref{eq:lcdmparameter}) with exact holding of the equality. Note that the values for the present comoving horizon $\eta_0$ also indicate that $\widetilde{r}$ is not important for the background expansion, while $\Omega_m$ does show its expected influence on $\eta_0$.
To see this, we focus on comparing the conditions of the models labeled with Nos.~(1),~(3),~(4),~(6), and (11), where different values of $\widetilde{r}$ rarely change $\eta_0$; on the 
other hand, a comparison among Nos.~(1),~(13), and (14) shows a slight dependence of $\eta_0$ on $\Omega_m$, as expected.
Especially, No.~(11) is a model extremely close to the $\Lambda$CDM model,
and the EoS of dark energy is almost constant $w_{\phi}\approx-1$, predicting a future evolution 
quickly approaching the de Sitter expansion.
}
\label{tab:para}
\end{center}
\end{table}


\subsection{Perturbations to light propagation and luminosity distance}
Following Refs.~\cite{FS1989,AOF2019}, as we have solved the metric perturbations $\Psi_\ell$ associated with large-scale fluctuations of the dark energy, we can evaluate the 
perturbation to the luminosity distance introduced by the inhomogeneities of the dark energy by considering the metric perturbations formulated previously.
The relative perturbations of the luminosity distance in an inhomogeneous universe is given as~\cite{FS1989,sasaki1987}
\begin{align}
 I &\equiv {\delta d_L \over d_L}
  =\int^{\lambda_s}_0 \diff \lambda {\lambda \over \lambda_s}(\lambda-\lambda_s) \left(\Delta^{(3)}\Psi-\left(\ddot{\Psi}+2{\diff\dot\Psi \over \diff\lambda}\right)\right),
  \label{eq:deltald}
\end{align}  
where $\dot\Psi\equiv
  {\partial \Psi(\eta,\chi) \over \partial \eta}$, and we have assumed a spatially flat universe. 
The traceless property of matrices $P_{ij}^{(m)}$ defined by 
Eq.~(\ref{def:Psi}) in $\Psi$ ensures that $\Delta^{(3)}\Psi=0$ [see Eq.~(\ref{laplacian})]. 

For the term containing differentiation with respect to the propagation parameter $\lambda$, we may write
\begin{align}
{\diff\over \diff\lambda}={\diff\eta\over \diff\lambda}{\partial \over \partial \eta}
+{\diff\chi\over \diff\lambda}{\partial \over \partial \chi}.
\label{eq:dlambda}
\end{align}
Here, we may take the parameter $\lambda$ as the comoving distance $\chi$; hence, $\lambda\equiv\chi=\eta_0-\eta$ and $\lambda_s\equiv\chi_s=\eta_0-\eta_s$ with an arbitrary light source indicated by subscript $s$. Thus, we have
\begin{align}
  I =& \int^{\chi_s}_{0} \diff \chi {\chi \over \chi_s}(\chi-\chi_s)
  \left(\ddot{\Psi}-2 {\partial \dot{\Psi} \over \partial \chi}\right).
\end{align}
Using a procedure similar to that used to transform Eq.~(\ref{eq:tempfluc1}) to Eq.~(\ref{eq:tempfluc2}),
with the definition of $\Psi$ in Eq.~(\ref{def:Psi}) and  Eqs.~(\ref{def:y1m})--(\ref{def:y2m}), we can rewrite $I$ as
\begin{align}
  I  =&
  \int^{\chi_s}_{0} \diff \chi(\chi-\chi_s) {\chi \over \chi_s}
  \left[
  \left(\ddot{\Psi}_{\ell(m)}-2 \dot{\Psi}_{\ell(m)} {\partial \over \partial \chi}\right)
  \left(\sum^{3}_{m=1} \chi Y_{\ell=1}^{(m)}(\theta,\varphi)+\sum_{m=1}^{5} \chi^2 Y_{\ell=2}^{(m)}(\theta,\varphi)\right)
  \right]
  \nonumber
  \\
  \equiv& \sum_{\ell=1}^2 \sum_{m=1}^{2\ell+1} 
  S_{\ell(m)} Y_\ell^{(m)}(\theta,\varphi),
  \label{eq:ld}
  \end{align}
  with the integral defined as
  \begin{align}
  S_{\ell(m)}\equiv\int^{\chi_s}_{0} \diff \chi {\chi-\chi_s\over \chi_s} \left(\chi^{\ell+1}\ddot{\Psi}_{\ell(m)}-2\ell\chi^\ell
  \dot{\Psi}_{\ell(m)}
  \right).
   \end{align}
It is worth reminding the reader again that $\Psi_{\ell(m)}(\eta)$ is only a function of $\eta$. 
$S_{\ell(m)}$ is the quantity that reflects the impact of accumulative corrections on the luminosity distance by the inhomogeneities of the dark energy, which can be evaluated numerically.

\begin{figure}[t]
    \includegraphics[width=0.47\linewidth]{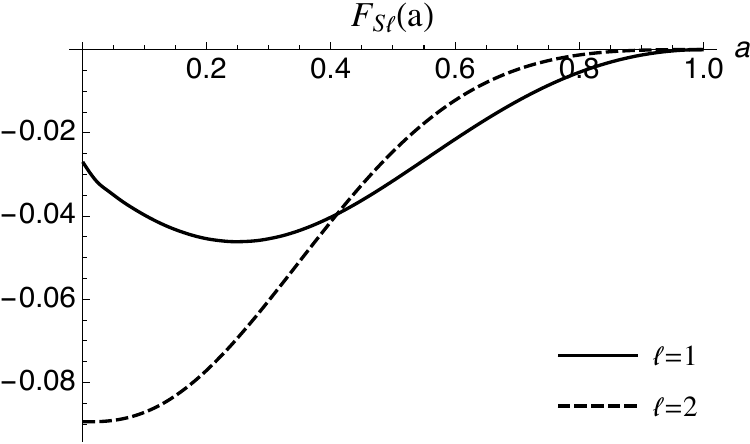}
\hspace{5mm}
    \includegraphics[width=0.47\linewidth]{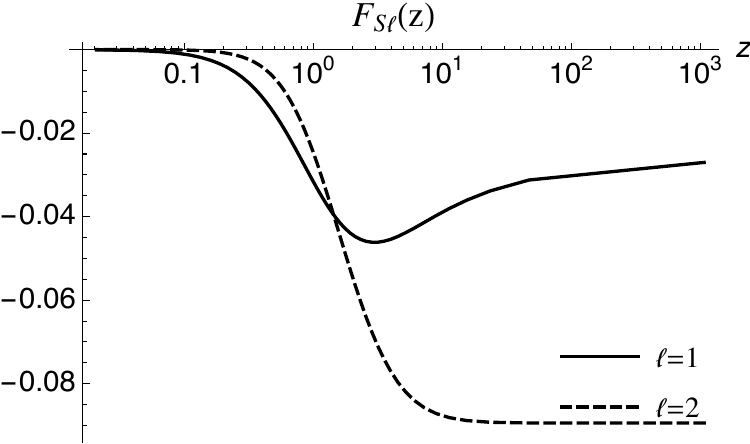}
\caption{Multipole components of the perturbations to the luminosity distance defined as $F_{S\ell(m)}(a;z)$ 
as a function of scale factor $a$ (left panel) and redshift $z$ (right panel). 
In each panel, the solid curve is the dipole, $\ell=1$, and the dashed curve is the quadrupole $\ell=2$. 
Here we adopted the model No.~(1) in the Table I.
}
\label{fig:fsaz} 
\end{figure}

We evaluate $S_{\ell(m)}$ due to the perturbation of $\Psi$ caused by dark energy inhomogeneity as a function of $a$ or the cosmological redshift $z$, corresponding to the light sources from different epochs, 
\begin{align}
  S_{\ell(m)}(a)=F_{S\ell(m)}(a)\D_{(\ell m)}.
    \label{eq:slma}
\end{align}
Then we have 
\begin{align}
F_{S\ell(m)}(a)
&\equiv 
  \int^{\eta_s(a)}_{\eta_0} \diff\eta \left((\eta_0-\eta)^{\ell+1}{\partial^2\Psi_{\ell(m)}\over\partial\eta^2} 
  -2\ell(\eta_0-\eta)^{\ell}{\partial\Psi_{\ell(m)}\over\partial\eta} 
  \right)
  {\eta-\eta_s(a) \over \eta_0-\eta_s(a)},
    \label{eq:fsaeta}
\end{align}
in a more explicit manner for numerical evaluation with respect to scale factor $a$ using $a_1$ as the variable of integration, 
\begin{align}
  F_{S\ell(m)}(a)
  &=
   -\int^{1}_{a} \diff a_1 
  \left[
  \left(\eta_0-\eta(a_1)\right)^{\ell+1}
  {\partial \over \partial a_1} \left(a_1^2 H(a_1) {\partial\Psi_{\ell(m)} \over \partial a_1} \right) 
  - 2\ell \left(\eta_0-\eta(a_1)\right)^{\ell}{\partial\Psi_{\ell(m)}\over\partial a_1} 
  \right]{\eta(a_1)-\eta_s(a) \over  \eta_0-\eta_s(a)}.
  \label{eq:fsaa}
\end{align} 
We notice that $F_{S\ell(m)}(a)$ does not increase or decrease monotonically, whose typical behaviors are illustrated as a function of $a$ or $z$ in Fig.~\ref{fig:fsaz}.
The scale factor is related to the cosmological redshift by $z=a^{-1}-1$, which is used to 
convert each other. 


Because we have solved for the system as functions of the scale factor $a$ in Sec.~\ref{sec:numer}, $\Psi_\ell(a)$, $\eta(a)$, and $H(a)$, and the particle horizon $\eta_0$ are already known for the given parameters $\widetilde{r}$ and $\widetilde{m}$ . If necessary, we can also transform these quantities using the conformal time $\eta$ 
as an independent variable (see Appendix~\ref{appen:transf}). On the other hand, we put constraints on $\varepsilon_1$ and $\varepsilon_2$ in Sec.~\ref{sec:appli_CMB}; 
thus, we can evaluate the modification to the luminosity distance $I$ with Eq.~(\ref{eq:ld}) by numerically evaluating $S_{\ell(m)}$ 
with the constraint Eqs.~(\ref{constr1})~and~(\ref{constr2}). 
Our numerical results $F_{S\ell(m)}$ are shown in Fig.~\ref{fig:fsaz} as a function of $a$ (left panel) and $z$ (right panel), respectively. 
Our results with different parameters can be found in Table~\ref{tab:para}. 
We evaluated $F_{S\ell(m)}({a})$ at $a=0.25$, which corresponds to $z=3$. 

We estimate the multipole components of $I$ as
\begin{align}
    I_\ell\equiv\sum_{m=1}^{2\ell+1} S_{\ell(m)}
 \sim(2\ell+1) S_{\ell(m)}.
\end{align}

Allowed values of $\D_{(
\ell m)}\sim\mathcal{O}(\varepsilon_\ell)$ ($\ell=1,2$) are found 
in Sec.~\ref{sec:appli_CMB} [see Eqs.~(\ref{constr1}) and (\ref{constr2})];
e.g., 
with $\varepsilon_1<1.2\times10^{-2}$ and $\varepsilon_2<5.7\times10^{-5}$, 
we can evaluate the modification to the luminosity distance caused by large-scale vector modes using Eq.~(\ref{eq:ld}) that the magnitude of the correction caused by the $\ell=1$ component is $\mathcal{O}(10^{-3})$, whereas it is $\mathcal{O}(10^{-5})$ for the $\ell=2$ component.
We have the consistent results of modification to the luminosity distance $I_\ell$ as
\begin{align}
  I_{\ell=1}
  \simeq-1.6\times10^{-3},
  \label{eq:estima1}
  \\
   I_{\ell=2}
  \simeq-2.0\times10^{-5},
 \label{eq:estima2}
\end{align}
at the redshift $z=3$ for all models in Table~\ref{tab:para}.


\section{Discussions and Conclusions}
\label{sec:discuss}

We formulated a cosmological model with inhomogeneous dark energy sourced from a dynamical scalar field with extremely large-scale fluctuations, 
by handling them as cosmological perturbations to a homogeneous background to focus on a local observable universe. 
This model is capable of reproducing an observable universe that mimics the $\Lambda$CDM flat universe favored by the observations but with inhomogeneity and anisotropy of small amplitudes on very large scales. 
We investigated the basic equations governing the evolution of the universe for the background and perturbations,
and presented the numerical solutions for these equations by choosing appropriate parameters that reproduce cosmological models close to the $\Lambda$CDM universe. 
As the examples for the application of the results, we investigated
the impact of the extremely large-scale inhomogeneity of the dark energy
on the cosmological observations in the late-time universe, where dark energy becomes important for background evolution. 

In our numerical evaluations, we chose the parameters of the models close to the $\Lambda$CDM model, for example, $(\widetilde{r}=70, \widetilde{m}=1/10)$ 
and $(\widetilde{r}=6.3, \widetilde{m}=1/3)$, which satisfies the condition in Eq.~(\ref{eq:lcdmparameter}). However, the prediction of 
the models is robust for the different choices of the parameters ($\widetilde{r}$, $\widetilde{m}$), as shown in Table~\ref{tab:para} in Sec.~\ref{sec:numer}. 
We also showed that slight changes in the values of $\Omega_m$ do not 
alter the results.
The observational constraints on cosmological parameters allow deviation from the standard $\Lambda$CDM scenario to some extent~\cite{TSJ,Jassal2010}, potentially suggesting that dynamical quintessence models for dark energy EoS are favored~\cite{DiValentino2020}.
Hence it is interesting 
to investigate constraints on the parameter space consistent with these observations.

Using numerical solutions, we focused our investigation on the impact of the large-scale inhomogeneities of the dark energy on the large angular anisotropies in the CMB temperature map and in the luminosity distance. 
The time variations of the metric perturbations give rise to the ISW effect, which affects the temperature anisotropies. 
In contrary to previous work \cite{scmde1}, we investigated the 
multipole spectrum in the spatially flat universe using numerical solutions without approximations. 
We obtained the constraints Eqs.~(\ref{constr1})~and~(\ref{constr2}) on the 
amplitude of the models from the observational data. 
The contribution from the large-scale inhomogeneities of the dark energy on the dipole of the CMB temperature power spectrum may partly account for 
the anomalies in the dipole and low multipoles of the CMB power spectra~\cite{Bielewicz2004,Polastri2015}.

The inhomogeneities of the dark energy affect the cosmic distance, which may 
impact the observations of SNe Ia and BAO measurements. 
We used Eq.~(\ref{eq:deltald}), according to Refs.~\cite{FS1989,AOF2019} for evaluation of these parts.
Our numerical calculations showed that 
the relative correction to the luminosity distance 
could be $\mathcal{O}(10^{-3})$ for the 
dipole and $\mathcal{O}(10^{-5})$ for the quadrupole components. 
For general parameter choices in Table~\ref{tab:para}, 
these corrections seem too small to resolve the Hubble tension, which is becoming increasingly conspicuous between measurements via CMB and via standard candles such as SNe Ia \cite{Nielsen2016,Mohayaee2020,Colin2019a}, as addressed in Sec.~\ref{sec:intro}. In a potentially related work Ref.~\cite{RGCai2021} as a comparison with our model, the authors attempted to ease the $H_0$ tension by introducing local inhomogeneities from the coupling of a chameleon dark energy model with dark matter.
However, comprehensive analyses, including wide ranges of the model parameters and the various observational results taking systematics into account, will be interesting ~\cite{Rubin2016,Rubin2020,DiValentino2020}.
Especially, the future progress of the gravitational wave observations with associative 
electromagnetic observations will be promising to provide with a standard siren \cite{Holz2005,Dalal2006,Vitale2018,Zhang2019}. 

Our model presented here is a possible dark energy model predicting the anisotropic expansion rate or anisotropic dark energy density and equation of state. 
Using the solutions in the present paper, we can realize the dynamical 
dark energy models with the inhomogeneous 
density on the large scales on the smooth background of the local universe. Another application of this inhomogeneous model may be to investigate its prediction on the structure formation, especially on large scales, though it is beyond the scope of this paper. Potentially related to this aspect, Refs.~\cite{Bagla1,Bagla2} investigated the impacts of different classes of dark energy models on matter clustering, which may help to discriminate our model from other models in the context of matter clustering.
The inhomogeneous dark energy model will be interesting from the viewpoint
that it 
is potentially verifiable/falsifiable
by the ongoing/planned data release of existing observations and future generation observations [for example, 
DES, DESI, LSST~\cite{LSST}, Euclid~\cite{Euclid}, and Roman Space Telescope (formerly known as WFIRST)~\cite{WFIRST}, cf.~\cite{Yamauchi2018}].
Additionally, the neutral hydrogen cosmology from the 21 cm spectrum survey planned by SKA~\cite{SKA} may link BAO with redshift-space distortions and add up to a better understanding of dark energy.
The future data of these surveys may help to test the
inhomogeneous properties of the dark energy.

The work in the present paper is inspired by a previous work \cite{scmde1},
in which large-scale dark energy perturbations are
generated by the quantum fluctuations of a scalar field 
according to an open-inflation scenario. The original model predicts a cosmological model with negative spatial curvature. However, in the present study, we considered a spatially flat universe  $\Omega_K=0$;
therefore, the origin of the scalar field as the candidate for dark energy  in our model is a subject to be discussed further.
Recently, ultralight scalar fields such as axionlike particles have attracted great interest as cosmological candidates for dark energy and dark matter~\cite{Visinelli:2018utg}, linked with the 
strong $CP$ problem and motivated by the string axiverse and the swampland conjectures ~\cite{Arvanitaki,Witten,Garg:2018reu,Ooguri,Heisenberg,Mizuno2019}.
Exploring the possibility to generate the initial conditions necessary for a scalar 
field in our model could be interesting within the framework of these scenarios in future investigations.

\acknowledgments

This work was supported by Grant-in-Aid for JSPS fellows Grant Number JP20J13640 (Y.N.), and MEXT/JSPS KAKENHI Grants No. 15H05895, No. 16H03977, No. 17K05444, No. 17H06359 (K.Y.).
We thank K. Yamashita, Y. Sugiyama, Y. Kojima, N. Okabe,
A. Naruko and M. Sasaki for fruitful discussions and helpful comments. We are also grateful for anonymous referees, whose comments improved the quality of the manuscript.

\appendix

\section{Multipole Expansion Matrices}
\label{appen:matrix}
The matrices appearing in the definitions of the perturbations in Sec.~\ref{sec:setup}, 
$P^{(m)}_{i}$ are simply written as 
\begin{align}
  P^{(m=1)}_{i} = \sqrt{3\over 4\pi}
    \left( 
      \begin{array}{ccc}
       1 \\
       0 \\
       0
      \end{array} 
      \right)
      \qquad \qquad
  P^{(m=2)}_{i} = \sqrt{3\over 4\pi}
    \left( 
      \begin{array}{ccc}
        0 \\
        1 \\
        0
       \end{array} 
      \right)
      \qquad \qquad
  P^{(m=3)}_{i} = \sqrt{3\over 4\pi}
    \left( 
     \begin{array}{ccc}
      0 \\
      0 \\
      1
     \end{array} 
    \right),
\end{align}  
while $P^{(m)}_{ij}$ are traceless matrices related to the multipole expansion of the perturbations and are listed as follows:
\begin{gather}
  P^{(m=1)}_{ij} =  \sqrt{15\over 16\pi} 
    \left( 
      \begin{array}{ccc}
       ~~0 & ~~1 & ~~0~~\\
       ~~1 & ~~0 & ~~0~~ \\
       ~~0 & ~~0 & ~~0~~
      \end{array} 
      \right),
      \\
  P^{(m=2)}_{ij} = \sqrt{15\over 16\pi} 
    \left( 
      \begin{array}{ccc}
        ~~0 & ~~0 & ~~0~~ \\
        ~~0 & ~~0 & ~~1~~ \\
        ~~0 & ~~1 & ~~0~~
      \end{array} 
      \right),
      \\
  P^{(m=3)}_{ij} = \sqrt{15\over 16\pi} 
    \left( 
      \begin{array}{ccc}
        ~~0 & ~~0 & ~~1~~ \\
        ~~0 & ~~0 & ~~0~~ \\
        ~~1 & ~~0 & ~~0~~
      \end{array} 
      \right),
      \\
  P^{(m=4)}_{ij} = \sqrt{15\over 16\pi}
    \left( 
      \begin{array}{ccc}
        ~~1 & ~~0 & ~~0~~ \\
        ~~0 & -1  & ~~0~~ \\
        ~~0 & ~~0 & ~~0~~
      \end{array} 
      \right),      
    \\
  P^{(m=5)}_{ij} = \sqrt{15\over 16\pi}
    \left( 
      \begin{array}{ccc}
        -1  & ~~0 & ~~0~~ \\
        ~~0 & -1  & ~~0~~ \\
        ~~0 & ~~0 & ~~2~~
      \end{array} 
      \right). 
\end{gather}

Equations~(\ref{def:Psi})---(\ref{def:field}) are due to the multipole expansion of the inhomogeneous perturbations under the real spherical harmonics: 
in the space up to $\ell=2$, the quadrupole component, with the $\ell=0$ component representing the homogeneous background as the monopole.

Using $\theta$ and $\varphi$ to denote the polar and azimuthal angles in the spherical coordinates, respectively, taking the spatial basis
\begin{align}
  &x^1=\chi \sin \theta \cos \varphi ,
  \nonumber\\
  &x^2=\chi \sin \theta \sin \varphi ,
  \nonumber\\
  &x^3=\chi \cos \theta ,
\end{align}  
the relation between these matrices and the spherical harmonics can be understood as
\begin{align}
  &Y_{\ell=1}^{(m)}(\theta,\varphi) \equiv P^{(m)}_{i}x^i /\chi,
  \label{def:y1m}
  \\
  &Y_{\ell=2}^{(m)}(\theta,\varphi) \equiv P^{(m)}_{ij} x^i x^j /\chi^2,
  \label{def:y2m}
\end{align}
with integer $m \in [1,2\ell+1]$ instead of $m \in [-\ell,\ell]$, corresponding to the three matrices for $\ell=1$ and five matrices for $\ell=2$ previously.

Note that the traceless property for the matrices corresponds to the conclusion that the large-scale modes make no 
source term contribution additional to the scalar modes as its Laplacian vanishes
\begin{align}
  \Delta^{(3)}\Psi=\nabla^2\Psi
  &=\Psi_{1(m)}\nabla^2 P_i^{(m)} x^i+\Psi_{2(m)}\nabla^2 P^{(m)}_{ij} x^i x^j
  \nonumber\\
  &=0+{\rm Tr}P^{(m)}_{ij} \Psi_{2(m)}\nabla^2 \chi^2
  \nonumber\\
  &=0.
  \label{laplacian}
\end{align}

\section{The Fluid Equation Consistency}
\label{appen:fluideq}
Starting from Eq.~(\ref{eq:flu}), 
if we refine ${\mathcal{V}}_\ell$ with respect to $V_\ell$ as
\begin{gather}
 \mathcal{V}_\ell \equiv{k \over a} V_\ell,
 \label{eq:ref_flu}
\end{gather}
with $k$ denoting the wave number of the perturbations, we will have 
\begin{align}
  \dot{V_\ell}-a \Psi_\ell= \frac{\dot a \mathcal{V}_\ell +a \dot{\mathcal{V}}_\ell}{k}- a \Psi_\ell=0,
\end{align}  
and hence 
\begin{align}
  \dot{\mathcal{V}}_\ell+ {\dot{a} \over a} \mathcal{V}_\ell-k\Psi_\ell=0,
\end{align}  
which is consistent with Eq.~(26) in Ref.~\cite{scmde1}.


\section{Additional Details for the Background Evolution\label{appen:background}}

In this appendix, we present more details for preparing the background equations for numerical solutions with the initial conditions.
\subsubsection{As functions of dimensionless time $\tilde{t}$}
\label{subsec:background-t}

We have introduced
the dimensionless ordinary differential equations of the background using $\tilde{t}$ as an independent variable as
\begin{align*}
   \widetilde r \widetilde{m}^2\widetilde{\phi}_0^2(\tilde{t})+\widetilde r \left(\frac{\diff\widetilde{\phi}_0}{\diff\tilde{t}} \right)^2+\Omega_m a^{-3}
   &=\left(\frac{1}{a}\frac{\diff a}{\diff\tilde{t}}\right)^2,
  \\
  \frac{\diff^2\widetilde{\phi}_0}{\diff\tilde{t}^2} + 3 \frac{1}{a} \frac{\diff a}{\diff\tilde{t}}
  \frac{\diff\widetilde{\phi}_0}{\diff\tilde{t}}+\widetilde{m}^2 \widetilde{\phi}_0
  &=0, 
\end{align*}
where $H_0$ is the Hubble constant, and 
$\bar\phi_0$ is a constant related to the initial  value of $\phi_0$.

It is worth noting that according to the definitions in Eqs.~(\ref{def:ndt})~to~(\ref{def:ndm}), there are 2 degrees of freedom for 
the parameters $\widetilde{m}$ and $\widetilde{r}$, to specify the mass and energy scale of the dark energy field $\phi$, respectively. The unknown component in our model, dark energy $\phi$, can be fundamentally characterized by two parameters. One is the shape of its potential $V(\phi)=\msq\phi^2/2$, and the other is the initial value
in our universe, while the properties of the other component (e.g., matter) are considered as known under the standard cosmological model.

To focus on the solution, in search of initial conditions and the analytic approximations in the limit $a \ll 1$, Eq.~(\ref{eq:tnd1}) approaches 
\begin{align}
  \left(\frac{1}{a}\frac{\diff a}{\diff\tilde{t}} \right)^2=\Omega_m a^{-3},
\label{eq:friedmd} 
\end{align}
which has the solution
\begin{align}
  \tilde{t} = \frac{2}{3} \frac{a^\frac{3}{2}}{\sqrt \Omega_m} 
  \qquad \textrm{or} \qquad 
  a= \left(\frac{9}{4}\Omega_m \right)^{\frac{1}{3}} \tilde{t}^{\frac{2}{3}}
  \label{eq:ini_a}
\end{align}
as an analytic approximation in the limit $a\ll1$.

Inserting this into Eq.~(\ref{eq:tnd2}) gives
\begin{align}
  \frac{\diff^2\widetilde{\phi}_0}{\diff\tilde{t}^2} + 2 \frac{1}{\tilde{t}}(\frac{\diff\widetilde{\phi}_0}{\diff\tilde{t}})+\widetilde{\msq}\widetilde{\phi}_0=0,
\end{align}
which has the general solution
\begin{align}
  \widetilde{\phi}_0(\tilde{t})= C_1 \frac{\sin(\widetilde{m}\tilde{t})}{\widetilde{m}\tilde{t}}+ C_2\frac{\cos(\widetilde{m}\tilde{t})}{\widetilde{m}\tilde{t}} .
\end{align}

The cosine part diverges in the limit $a \ll 1$ to be abandoned; hence, we write
\begin{align}
    \widetilde{\phi}_0(\tilde{t})= C_1 \frac{\sin(\widetilde{m}\tilde{t})}{\widetilde{m}\tilde{t}},
    \label{eq:iniphi}
\end{align}  
which imposes the initial condition
\begin{align}
\widetilde{\phi}_0(\tilde{t}\to 0)= \lim_{\tilde{t} \to 0}C_1\frac{\sin(\widetilde{m} \tilde{t})}{\widetilde{m} \tilde{t}} =C_1.
 \label{eq:bc_choice}
\end{align}
However, the initial value of  $\widetilde\phi_0(\tilde{t}\to0)=C_1$ is not self-evident and should be determined in association with the dark energy density of the present epoch inferred from observations.
We have the constraint from the present Hubble rate to fix $\tilde{t}_0$ by definitions
\begin{align*}
   a(\tilde{t}_0)= a(H_0t_0) &\equiv 1,
  \\
   H(\tilde{t}_0) = H(H_0{t_0}) &\equiv H_0.
\end{align*}
Inserting this into Eq.~(\ref{eq:tnd1}) actually gives Eq.~(\ref{eq:constr}),
\begin{align*}
  1-\Omega_m= \widetilde{r} \widetilde{m}^2 \left(\widetilde{\phi}_0\Big|_{\tilde{t}
  =\tilde{t}_0}\right)^2+\widetilde{r}\left(\frac{\diff \widetilde{\phi}_0}{\diff \tilde{t}}\bigg|_{\tilde{t}=\tilde{t}_0}\right)^2.
\end{align*}
Equation~(\ref{eq:constr}) is the necessary condition for specifying the dark energy density observed today when solving the background equations.
Together with Eqs.~(\ref{eq:tnd1}) and (\ref{eq:tnd2}), the system is now prepared for numerical evaluation to obtain the evolution of $a(\tilde{t})$ and $\widetilde{\phi}_0(\tilde{t})$. 
As we are mainly interested in the late-time evolution here, 
we can determine the initial value for independent variables $\tilde{t}$ or $a$ (to be discussed later) manually as a typical value; for example, $a_i = a_d \approx 1/1100$ at the photon decoupling off the last scattering,
by use of Eq.~(\ref{eq:ini_a}). These solutions determine the background evolution that we rely on to solve  the perturbation equations.

It is worth mentioning that Eq.~(\ref{eq:constr}) also provides a baseline for choosing the parameters $\widetilde{m}$ and $\widetilde{r}$ from the various parameter spaces. 
In the case of the cosmological constant $\Lambda$, $\diff \widetilde{\phi}_0/\diff \tilde{t}$ is always small, leaving
\begin{align}
  1-\Omega_m= \widetilde{r} \widetilde{m}^2 \left(\widetilde{\phi}_0\Big|_{\tilde{t}=\tilde{t}_0}\right)^2.
\end{align}
Thus, if we take the dimensionless field in the present epoch normalized as $\widetilde{\phi}_0(\tilde{t}=\tilde{t}_0)\sim\mathcal{O}(1)\equiv1$, we will have a special 
case for the choice of parameters approximating the $\Lambda$CDM model presented in Eq.~(\ref{eq:lcdmparameter}) that $\widetilde{r} \widetilde{m}^2 \simeq 1-\Omega_m$.

\subsubsection{As functions of scale factor $a$}
\label{subsec:background-a}
Because the scale factor $a$ can be chosen as a time-evolution parameter instead of the dimensionless time $\tilde{t}$, 
as a double-check for the previous subsection, we can write out the dimensionless equations for $\widetilde{\phi}_0$ and $\widetilde{H}(a)$ as functions of the scale factor $a$. Recall that the superscript $'$ means derivative with respect to scale factor $a$, 
By inserting Eq.~(\ref{eq:ha2}) into Eq.~(\ref{eq:phia2}), we obtain the background equation to be solved in the form
\begin{align}
  \widetilde{m}^2 a^2 \widetilde{\phi}_0 (1-\widetilde{r}a^2\widetilde{\phi}_0'^2)
    +\widetilde{m}^2 \widetilde{r} a^3 \widetilde{\phi}_0^2 \left(4\widetilde{\phi}_0'-3\widetilde{r}a^2\widetilde{\phi}_0'^3+a\widetilde{\phi}_0'' \right)
    +{\Omega_m \over 2}\left(5\widetilde{\phi}_0'-3\widetilde{r}a^2\widetilde{\phi}_0'^3+2a\widetilde{\phi}_0'' \right)
    =0,
  \label{eq:phia3}
\end{align}
where an initial condition for $\widetilde{\phi}_0(a)$ is necessary. After solving $\widetilde{\phi}_0(a)$, we can obtain  $\widetilde{H}(a)$ from Eq.~(\ref{eq:ha2}).

For the initial conditions, we consider the analytic approximations.  When $a \ll 1$,  Eq.~(\ref{eq:ha2}) simply approaches 
  \begin{align}
    \widetilde{H}=\sqrt{\Omega_m}a^{-3/2}.
  \end{align}
Inserting this into Eq.~(\ref{eq:phia2}) and simplifying will lead to
  \begin{align}
    a \widetilde{\phi}_0''+{5 \over 2}\widetilde{\phi}_0'+\widetilde{m}^2 a^2 \Omega_m^{-1} \widetilde{\phi}_0=0,
  \end{align}
which can be solved analytically as 
  \begin{align}
    \widetilde{\phi}_0(a)= C_1 \frac{3 \sqrt{\Omega _m} }{2 \widetilde{m} a^{3/2} } \sin \left(\frac{2 \widetilde{m} a^{3/2} }{3 \sqrt{\Omega _m}}\right) ,
  \end{align}
which is identical to that in Eq.~(\ref{eq:iniphi}) by recalling Eq.~(\ref{eq:ini_a}). 
Then we are able to infer 
  \begin{align}
    \widetilde{\phi}_0(a\rightarrow0)&=C_1,
    \\
   \widetilde{\phi}_0'(a\rightarrow0)&=0,
  \end{align}
are the appropriate initial conditions for the system, which are consistent with the equations using dimensionless time $\tilde{t}$ as the independent variable.

Now the background equations can be solved numerically. 

\section{Dark energy EoS as a function of the scale factor $a$}
\label{appen:EOSCPL}
Following Eq.~(\ref{def:eos}), we have 
\begin{align}
  \omega_\phi 
  &=-{\msq a^2\phi^2-\dot\phi^2 \over \msq a^2\phi^2+\dot\phi^2}
  \nonumber\\
  &=-1+{2 \over \msq a^2 (\phi/\dot\phi)^2+1}
  \nonumber\\
  &\equiv-1+2 W(a),
  \label{eq:eoscpl2}
\end{align}
where $'$ denotes the derivative of the scale factor $a$, ${}'\equiv\partial/\partial a$, where
\begin{align}
  W(a)\equiv{1 \over \msq a^2 (\phi/\dot\phi)^2+1}={a^2 \over (\mp/H)^2(\phi/\phi')^2+a^2}.
  \label{eq:eosparam1}
\end{align}  
At the background level assuming $\widetilde{\phi}\simeq\widetilde{\phi}_0$, we can further write
\begin{align}
  W(a)\simeq{\widetilde\phi_0'^2 a^2 \widetilde{H}^2 \over \widetilde{m}^2 \widetilde\phi_0^2+\widetilde\phi_0'^2 a^2 \widetilde{H}^2}.
  \label{eq:eosparam2}
\end{align}
Recall that $\widetilde{H}(a)$ is defined in Eq.~(\ref{eq:ha2}), which depends on the values of $\widetilde{r}$, $\widetilde{m}$, and  $\Omega_m$. Because $\widetilde{m}^2$ is typically small in our model, using Eq.~(\ref{eq:ha2}) and expanding to the order of $\mathcal{O}(\widetilde{m}^2)$,  we have 
  \begin{align}
      W(a)
      &\simeq
      1-\frac{a\widetilde{m}^2}{\Omega_m}
      \left(\frac{\widetilde{\phi}_0^2}{\widetilde{\phi}_0'^2}\right)
      \left(1-\widetilde{r}a^2\widetilde{\phi}_0'^2\right)
      \nonumber
      \\
      &\simeq
      1-\frac{a\widetilde{m}^2}{\Omega_m}
      \left(\frac{\widetilde{\phi}_0^2(a)}{\widetilde{\phi}_0'^2 (a)}\right),
      \label{eq:eosparam3}
  \end{align}
which can be numerically evaluated with 
$\widetilde\phi_0(a)$ and $\widetilde\phi_0'(a)$ as demonstrated 
in Sec.~\ref{sec:numer}. 
The second line stands because $a^2\widetilde{\phi}_0'^2$ is small and negligible 
for $0<a<1$ in almost all cases. Then, we can understand that 
$\widetilde{r}$ hardly affects the background EoS of the dark energy. 

Although slightly complicated in its explicit form, Eqs.~(\ref{eq:eoscpl2})--(\ref{eq:eosparam3}) can be considered as a natural extension of the CPL parametrization of the dark energy EoS \cite{ChePolar,Linde0}. 
This is a manifestation of how the behaviors of the EoS of dark energy 
in our model are decided quantitatively by the parameters.

If we include the first-order perturbations 
$\widetilde\phi_\ell(a)$ in Eq.~(\ref{eq:eoscpl2}), and hence corrections to (\ref{eq:eosparam3}), we can evaluate the anisotropies of the EoS $w_{\phi}(a)$ of dark energy 
sourced by the inhomogeneities of $\phi$,
although these corrections to the isotropic background in Eq.~(\ref{eq:eosparam3}) may be small because of our previous constraints on the amplitudes of $\varepsilon_1$ and $\varepsilon_2$ 
in Eqs.~(\ref{constr1}) and (\ref{constr2}).

\section{Luminosity Distance}
\label{appen:ld}

Starting from Eq.~(\ref{eq:deltald}),  while still taking the propagation parameter as $\lambda=\chi$, including the normal mode scalar fluctuation $\Psi^{\rm tot}=\Psi^{\rm norm}+\Psi$, to the linear order, we obtain
\begin{align}
  I^{\rm tot}_{\rm lin}=\int^{\chi_s}_0 \diff \chi {\chi \over \chi_s} (\chi-\chi_s)\Delta^{(3)}\Psi^{\rm tot};
\end{align} 
but we have Eq.~(\ref{laplacian}) for $\Psi$; hence, we only need to consider the cosmological Poisson equation as
\begin{align}
  \Delta^{(3)}\Psi^{\rm tot}=\Delta^{(3)}\Psi^{\rm norm}=4\pi G \bar\rho_m \delta_m a^2.
\end{align} 
In gravitationally bound local systems, for example, where objects such as SNe Ia are located, the source term of the scalar perturbations from matter in the Friedmann equation simply reads
\begin{align}
 8\pi G \bar\rho_m =3H^2=3H_0^2 \Omega_m a^{-3},
\end{align} 
which is identical to Eq.~(\ref{eq:rho0});
hence,  
\begin{align}
  I^{\rm tot}_{\rm lin}=-{3H_0^2\Omega_m \over 2}\int^{\chi_s}_0 \diff \chi {\chi \over \chi_s} (\chi_s-\chi)a(\chi)^{-1}\delta_m(a(\chi),{\bm\gamma}).
\end{align} 
Because $a^{-1}=1+z$ holds by definition between scale factor $a$ and cosmological redshift $z$, if we only look at the contribution by the inhomogeneous background and neglect
peculiar motion terms, this result is consistent with Eq. (6) in Ref.~\cite{AOF2019}.

\section{Some Useful Transformation Relations}
\label{appen:transf}
Here, we provide some useful relations to help transform equations quickly between forms as functions of $\tilde{t}$, $a$, or $\eta$. 
As we defined the dimensionless quantities in Eqs.~(\ref{def:ndt}) and (\ref{def:ndh}),
\begin{gather}
 \tilde{t}=H_0 t,
 \nonumber
 \\
 \widetilde{H}=H/H_0,
 \nonumber
\end{gather}
with 
\begin{gather}
    H={1\over a}{\diff a \over \diff t}
\end{gather}
as a usual convention.
Hence, recalling $'$ is the derivative with respect to $a$ and overdot $\dot{\quad}$ indicates that with respect to $\eta$, 
for arbitrary function $\mathcal{A}$ we have
\begin{gather}
    {\partial \mathcal{A} \over \partial \tilde{t}}={\partial \mathcal{A}\over H_0\partial t}=a{H\over H_0}{\partial \mathcal{A} \over \partial a}=a\widetilde{H}\mathcal{A}',
    \label{eq:trans-ndt-a}
\end{gather}
as well as
\begin{gather}
  {\partial \mathcal{A} \over \partial \tilde{t}} = {\partial \mathcal{A}\over H_0\partial t}=\frac{1}{a H_0}{\partial \mathcal{A}\over \partial \eta}=\frac{\widetilde{H}}{a H}{\partial \mathcal{A}\over \partial \eta}=\frac{\widetilde{H}}{\mathcal{H}}\dot{\mathcal{A}}.
  \label{eq:trans-ndt-eta}
\end{gather}  
These will help to transform equations quickly.
Following these we have 
\begin{align}
  {\partial^2\mathcal{A} \over \partial \tilde{t}^2}
  &=a\widetilde{H} {\partial \over \partial a} \left(a \widetilde{H} {\partial \mathcal{A} \over \partial a} \right) \nonumber
  \\
  &=a^2\widetilde{H}^2\mathcal{A}''+(a^2\widetilde{H}\widetilde{H}'+a\widetilde{H}^2)\mathcal{A}'
\end{align}  
and 
\begin{gather}
    {1\over a}{\partial a \over \partial \tilde{t}}={1\over H_0}{1\over a}{\partial a \over \partial t}=H/H_0=\widetilde{H},
\end{gather}
as very useful relations.

Finally let us note a universal relation widely used,
\begin{gather}
    \dot{\mathcal{A}}=a^2 H \mathcal{A}'.
    \label{eq:trans-a-eta}
\end{gather}


\end{document}